%% file: main.tex
\documentclass[fleqn, usenatbib, useAMS]{mnras}
\input{header.tex}
\title[Fundamental scales in the turbulent dynamo]{Fundamental scales in the kinematic phase of the turbulent dynamo}
\author[Kriel, et al., 2022]{
    Neco Kriel$^{\orcidicon{0000-0002-3558-3926}\, 1}$\thanks{neco.kriel@anu.edu.au}, 
    James R.~Beattie$^{\orcidicon{0000-0001-9199-7771}\, 1}$, %\thanks{james.beattie@anu.edu.au}, 
    Amit Seta$^{\orcidicon{0000-0001-9708-0286}\, 1}$, %\thanks{amit.seta@anu.edu.au}, 
    and Christoph Federrath$^{\orcidicon{0000-0002-0706-2306}\, 1, 2}$ \\ %\thanks{christoph.federrath@anu.edu.au} \\
    $^{1}$Research School of Astronomy and Astrophysics, Australian National University, Canberra, Australia \\
    $^{2}$Australian Research Council Centre of Excellence in All Sky Astrophysics (ASTRO3D), Canberra, ACT 2611, Australia \\
}
\date{Accepted XXX. Received YYY; in original form ZZZ.}
\pubyear{2022}

\begin{document}

%% TITLE
\label{firstpage}
\pagerange{\pageref{firstpage}--\pageref{lastpage}}
\maketitle

%% ABSTRACT
\begin{abstract}
    The turbulent dynamo is a powerful mechanism that converts turbulent kinetic energy to magnetic energy. A key question regarding the magnetic field amplification by turbulence, is, on what scale, $\kp$, do magnetic fields become most concentrated? There has been some disagreement about whether $\kp$ is controlled by the viscous scale, $\knu$ (where turbulent kinetic energy dissipates), or the resistive scale, $\keta$ (where magnetic fields dissipate). Here we use direct numerical simulations of magnetohydrodynamic turbulence to measure characteristic scales in the kinematic phase of the turbulent dynamo. We run $104$-simulations with hydrodynamic Reynolds numbers of \mbox{$10 \leq \Reyk \leq 3600$}, and magnetic Reynolds numbers of \mbox{$270 \leq \Reym \leq 4000$}, to explore the dependence of $\kp$ on $\knu$ and $\keta$. Using physically motivated models for the kinetic and magnetic energy spectra, we measure $\knu$, $\keta$ and $\kp$, making sure that the obtained scales are numerically converged. We determine the overall dissipation scale relations \mbox{$\knu = (0.025^{+0.005}_{-0.006})\, k_\turb\, \Reyk^{3/4}$} and \mbox{$\keta = (0.88^{+0.21}_{-0.23})\, \knu\, \Pranm^{1/2}$}, where $k_\turb$ is the turbulence driving wavenumber and $\Pranm=\Reym/\Reyk$ is the magnetic Prandtl number. We demonstrate that the principle dependence of $\kp$ is on $\keta$. For plasmas where $\Reyk \gtrsim 100$, we find that \mbox{$\kp = (1.2_{-0.2}^{+0.2})\, \keta$}, with the proportionality constant related to the power-law `Kazantsev' exponent of the magnetic power spectrum. Throughout this study, we find a dichotomy in the fundamental properties of the dynamo where $\Reyk > 100$, compared to $\Reyk < 100$. We report a minimum critical hydrodynamic Reynolds number, $\Reyk_\crit = 100$ for bonafide turbulent dynamo action.
\end{abstract}

%% KEYWORDS

%% INTRODUCTION
\section{Introduction}\label{sec:theory}

The Universe is observed to be in a magnetised state on most astrophysical scales probed so far. This ranges from small scales, like planets \citep{stevenson2010planetary, jones2011planetary, sheyko2016magnetic} and stars \citep{choudhuri2015nature, beck2018magnetic}, over the interstellar medium of galaxies \citep{beck2001galactic, fletcher2011magnetic, han2017observing}, through to scales occupied by the largest gravitationally bound structures in the Universe, galaxy clusters \citep{clarke2001new, brandenburg2005astrophysical, vazza2014amplification, marinacci2018first}. Observing and probing these magnetic fields provide us with insight into the structure and dynamics of the astrophysical systems that they occupy, and the dynamically important roles that magnetic fields play \citep{subramanian2016origin, federrath2016magnetic, rincon2019dynamo, krumholz2019role}.

In the early Universe, magnetic fields were orders of magnitude weaker than they are observed to be in the present day, as most seed field generation mechanisms produce very weak magnetic fields \citep{grasso2001magnetic, schleicher2010small, widrow2012first, durrer2013cosmological}. While the origin of magnetic fields is still an unsolved problem, many theories predict weak magnetic fields of $\ll \n\G$ could be generated during various phases in the early Universe \citep[see][and references therein]{subramanian2016origin}. These magnetic fields are subject to resistive and turbulent decay, and in spiral galaxies they are removed via other processes, such as galactic winds and flux expulsion \citep{weiss1966expulsion, gilbert2016flux, Seta2019}. To explain the observed magnetic fields of the present day, weak primordial magnetic fields therefore need to amplify on a time-scale much faster than their dissipation and/or removal. \citet{wagstaff2014magnetic} showed that the conditions for an efficient ``turbulent dynamo'' (also referred to as a small-scale dynamo in literature) to act in the early Universe are satisfied. The turbulent dynamo is the only mechanism capable of providing exponentially fast amplification of magnetic fields, explaining the $\mu\mathrm{G}$ fields observed in galaxies today \citep{widrow2012first, beck2018magnetic, subramanian2019primordial}. The main aim of this study is to determine on what scale magnetic fields become most concentrated, and how this magnetic peak scale ($\kp$) relates to the viscous ($\knu$) and resistive ($\keta$) dissipation scales in the kinematic phase of the dynamo \citep{batchelor1950spontaneous, kazantsev1968enhancement, kulsrud1992spectrum, Vainshtein1992, schekochihin2002spectra, schekochihin2004simulations, brandenburg2005astrophysical, schober2015saturation, xu2016turbulent, mckee2020magnetic}.

\subsection{A hierarchy of scales}

Turbulence is associated with large hydrodynamic Reynolds numbers, 
\begin{equation}
    \Reyk \equiv \frac{\ell_\turb\, u_\turb}{\nu} , \label{eqn:Reyk}
\end{equation}
where $u_\turb$ is the flow velocity on the driving scale $\ell_\turb$, and $\nu$ is the kinematic viscosity. The $\Reyk$ provides a measure of the ratio of inertial to viscous forces in a gas. At low $\Reyk$, the viscous forces are dominant, and flows are laminar. Conversely, high $\Reyk$ flows are turbulent. The $\Reyk$ at which a flow transitions from laminar to turbulent depends on the geometry of the system, but fully developed turbulence generally develops for $\Reyk \gtrsim 100$--$1000$ \citep{frisch1995turbulence, schumacher2014small}.

For incompressible, homogeneous and isotropic turbulence, energy is transported from larger scales (where the turbulence is driven) to smaller scales (where the turbulent energy is dissipated by viscous forces). Eddies on length scale $\ell$ rotate with velocity $u_\ell \propto \ell/t_\ell \propto \ell^{1/3}$ \citep{Kolmogorov1941, frisch1995turbulence}, where $t_\ell \propto \ell / u_\ell$ is the scale-dependent turnover time. Thus, the viscous scale for such a turbulent flow, $\ell_\nu \propto u_\nu^3$, is the scale where dissipation effects are dominant, and can be defined by the condition that the Reynolds number on $\ell_\nu$, $\Reyk_{\ell_{\nu}}$ is unity, i.e., $\Reyk_{\ell_{\nu}} = \ell_\nu u_\nu / \nu \propto \ell_\nu^{4/3} \approx 1$. Thus, $\ell_\nu/\ell_\turb \propto \Reyk^{3/4}$, as per \autoref{eqn:Reyk}, and the wavenumber associated with the viscous scale eddies is
\begin{equation}
    \knu = \frac{2\pi}{\ell_\nu} \propto \knutheory \equiv k_\turb \, \Reyk^{3/4} , 
    \label{eqn:knutheory}
\end{equation}
where the turbulence driving wavenumber is $k_\turb = 2\pi / \ell_\turb$. Here, we use the symbol $\propto$ to emphasise that $\knu$ scales with $\knutheory$, but may not be equal to it. In fact, one of our goals in this work is to determine $\knu$ and whether the theoretical dependence on $\knutheory \equiv k_\turb \, \Reyk^{3/4}$ holds.

When magnetic fields are present, one can define the magnetic Reynolds number, 
\begin{equation}
    \Reym \equiv \frac{\ell_\turb\, u_\turb}{\eta} , \label{eqn:Reym}
\end{equation}
in analogy to the hydrodynamic Reynolds number (see \autoref{eqn:Reyk}), where $\nu$ is replaced by the magnetic resistivity, $\eta$, and $\Reym$ is a measure of the ratio between induction forces and magnetic dissipation. The relative importance between $\nu$ and $\eta$ can be quantified by the magnetic Prandtl number, 
\begin{equation}
    \Pranm \equiv \frac{\Reym}{\Reyk} = \frac{\nu}{\eta} . \label{eqn:Pranm}
\end{equation}
The $\Pranm$ controls the scale separation between $\ell_\nu$ and the resistive scale, $\ell_\eta$, and is an important parameter for characterising the behaviour and evolution of the magnetic energy in a turbulent plasma. In the $\Pranm \gg 1$ regime, \citet{schekochihin2002spectra} provide an estimate for the resistive wavenumber as
\begin{equation}
    \keta = \frac{2\pi}{\ell_\eta} \propto \ketatheory \equiv \knutheory \, \Pranm^{1/2}. \label{eqn:ketatheory}
\end{equation}
Here, we again use $\propto$ to emphasise that there may be a constant of proportionality between $\keta$ and $\ketatheory$, to be determined in this study. This leads to a hierarchy of scales in the $\Pranm > 1$ regime, with $\ell_\turb > \ell_\nu > \ell_\eta$. Here, $\ell_\turb > \ell > \ell_\nu$ defines the inertial range of Kolmogorov turbulence, and $\ell_\nu > \ell > \ell_\eta$ defines the sub-viscous range.

Generally, in most astrophysical systems, $\Reyk$ and $\Reym$ are large, and therefore both velocity and magnetic fields are expected to be turbulent and span over a wide range of scales. However, while magnetic fields in astrophysical settings, like in the Milky Way, could have a wide range of scales available, it is of interest to know the scale, $\kp$, on which these fields are expected to become most concentrated. \citet{batchelor1950spontaneous} argued that $\kp \propto \knu$, but more recent theories suggest that $\kp \propto \keta$ \citep{kazantsev1968enhancement, kulsrud1992spectrum, Vainshtein1992, schekochihin2002spectra, schekochihin2004simulations, brandenburg2005astrophysical, schober2015saturation, xu2016turbulent, mckee2020magnetic}. Our primary focus in this study is to determine the dissipation scales $\knu$ and $\keta$, and the magnetic peak scale $\kp$, and to compare our measurements with the theoretical predictions. We do so by utilising a suite of simulations of turbulent dynamo amplification, which span a large range of $\Reyk$ and $\Reym$. In all of our simulations, we measure $\knu$, $\keta$, and $\kp$, to test the theoretical relations given by \autoref{eqn:knutheory} and \ref{eqn:ketatheory}, and to determine the exact dependence of $\kp$ on $\knu$ and $\keta$.

\subsection{Magnetic field amplification}

While the theory of turbulent magnetic field amplification dates back to \citet{batchelor1950spontaneous} and \citet{kazantsev1968enhancement}, and to many follow-up works \citep[e.g., ][]{brandenburg2005astrophysical, federrath2016magnetic, rincon2019dynamo}, it is only within the last few years, that laboratory experiments have demonstrated that magnetic fields can be amplified by turbulent dynamo action \citep{meinecke2015developed, tzeferacos2018laboratory, bott2021time}. If initially the strength of the magnetic energy density, $E_\tmag = B^2 / (8\pi)$, is much weaker than the turbulent kinetic energy density, $E_\tkin = \rho_0 \, u_\turb^2 / 2$ (where $\rho_0$ is the mean gas density), then the turbulence is able to rapidly amplify the magnetic field \citep{batchelor1950spontaneous, kazantsev1968enhancement, kulsrud1992spectrum, Vainshtein1992, schekochihin2002spectra, schekochihin2004simulations, brandenburg2005astrophysical, schober2015saturation, xu2016turbulent, SetaEA2020sat, mckee2020magnetic, seta2021saturation} by randomly stretching, twisting, and folding the magnetic field lines \citep{vainshtein1972origin, zel1984kinematic, schekochihin2002spectra, schekochihin2004simulations, SetaEA2015stf}. This is called the kinematic phase of the dynamo, with the condition that $\Reym$ exceeds a critical value, $\Reym_\crit\gtrsim100$, depending on $\Pranm$ and the level of compressibility (sonic Mach number) of the plasma \citep{schekochihin2004critical, schekochihin2004simulations, haugen2004suppression, brandenburg2005astrophysical, schober2012magnetic, federrath2014turbulent}. The growth rate of the magnetic field has been shown to depend upon $\Reyk$ and $\Pranm$, with faster amplification associated with higher $\Reyk$ and $\Pranm$ \citep{subramanian1997dynamics, schober2012magnetic, schober2015saturation, bovino2013turbulent, federrath2014turbulent}. Once the magnetic field becomes strong enough such that the Lorentz force exerts a significant back-reaction on the turbulent flow, the stretching motions (which amplify magnetic fields) are ultimately suppressed, the diffusion relative to stretching is enhanced, and both these processes combined lead to the saturation (saturated phase) of the turbulent dynamo \citep{schekochihin2002model, SetaEA2020sat, seta2021saturation}.

The rest of the study is organised as follows. In \S\ref{sec:methods} we introduce our numerical methods and simulation parameters. In \S\ref{sec:results} we present the results of this work, starting with the magnetic-to-turbulent kinetic energy ratio in \S\ref{subsec:B_evolution}. In \S\ref{subsec:slices} we investigate the morphology of the kinetic and magnetic energy during the kinematic phase of the dynamo. In \S\ref{subsec:spectra_fits} we analyse the velocity and magnetic field power spectra, and introduce our models and methods for measuring $\knu$, $\keta$, and $\kp$ from the spectra. In \S\ref{subsec:relation} we compare where we measure $\knu$ and $\keta$ in our simulations with where theories predict these scales to be. In \S\ref{subsec:dependence} and \S\ref{subsec:kaz_exp} we determine the dependence of $\kp$ on the dissipation scales and link $\kp$ to the slope of the magnetic field spectrum. In \S\ref{subsec:limitations} and \S\ref{subsec:implications} we discuss the limitations and implications of the results within this study, respectively. Finally, we summarise this study and our results in \S\ref{sec:summary}.

%% NUMERICAL METHODS
\section{Numerical simulations} \label{sec:methods}

\subsection{MHD equations and numerical methods} \label{subsec:MHD_eqns}

We solve the compressible, non-ideal, magnetohydrodynamic (MHD) equations, 
\begin{align}
    \priv{\rho}{t} + \nabla\cdot(\rho\bm{u}) &= 0 , 
        \label{MHD:continuity} \\
    \rho\rbrac{\priv{}{t} + \bm{u}\cdot\nabla} \bm{u}
        &= \frac{1}{4\pi} (\bm{B}\cdot\nabla)\bm{B} - \nabla\rbrac{p_\text{th} + \frac{B^2}{8\pi}} \nonumber\\
        &\nquad[3] + \nabla\cdot\rbrac{2\nu\rho\bm{\mathcal{S}}} + \rho\bm{F} , 
        \label{MHD:momentum} \\
    \priv{\bm{B}}{t} 
        &= \nabla\times(\bm{u}\times\bm{B}) + \eta\nabla^2\bm{B} , 
        \label{MHD:induction} \\
    \nabla\cdot\bm{B} &= 0 , 
        \label{MHD:div_free}
\end{align}
for an isothermal plasma with constant kinematic viscosity, $\nu$, and magnetic resistivity, $\eta$. In these equations, $\rho$ is the gas density, $\bm{u}$ is the gas velocity, and $\bm{B} = \bm{B}_0 + \delta\bm{B}$ is the total magnetic field, which consists of a mean field, $\bm{B}_0$, (which we initialise as zero) and fluctuating, $\delta\bm{B}$, component (see \S\ref{subsec:initalise} for details on how we initialise $\delta\bm{B}$). The viscous dissipation rate is included in the momentum equation (\autoref{MHD:momentum}) via the strain rate tensor $\mathcal{S}_{ij} = (1/2)~(\pp_i u_j + \pp_j u_i) + (2/3)~\delta_{ij}\nabla\cdot\bm{u}$, where $\delta_{ij}$ is the Kronecker delta. $\bm{F}$ is the turbulent acceleration field, which we discuss in \S\ref{subsec:turb_driving}. We close the energy equation with an isothermal equation of state, $p_\text{th} = c_s^2 \rho$, where $p_\text{th}$ is the thermal pressure, and $c_s=\mathrm{const}$ is the sound speed. 

We use a modified version of the \textsc{flash} code \citep{fryxell2000flash, Dubey2008} to solve the MHD equations (\autoref{MHD:continuity}--\ref{MHD:div_free}) on a uniformly discretised, triply periodic, three-dimensional grid with dimensions $L^3$. We test numerical convergence by running our simulations with different grid resolutions, with up to $576^3$ grid cells (see \S\ref{subsec:resolution} below). For solving the MHD equations, we use the five-wave, approximate Riemann solver described in \citet{bouchut2007multiwave, Bouchut2010}, and implemented and tested in \citet{waagan2011robust}.

\subsection{Turbulence driving} \label{subsec:turb_driving}

The turbulent acceleration field, $\bm{F}$, in the momentum \autoref{MHD:momentum} is modelled with an Ornstein-Uhlenbeck process \citep{eswaran1988examination, schmidt2009numerical, federrath2010comparing}. We use the Helmholtz decomposition of $\bm{F}$ to control the solenoidal (divergence-free) and compressive (curl-free) modes in $\bm{F}$ \citep{federrath2008density, federrath2010comparing}. Here, we choose to drive with only solenoidal modes, because solenoidal driving is the most efficient at amplifying magnetic fields \citep{federrath2011mach, federrath2014turbulent, martins2019kazantsev, chirakkara2021efficient}. We choose to work in the subsonic, near incompressible regime of turbulence, with $\Pranm \geq 1$, because the turbulent dynamo is most efficient in this regime \citep[e.g., ][]{schekochihin2004simulations, schekochihin2007fluctuation, schober2012magnetic, schober2015saturation, seta2020seed, chirakkara2021efficient, seta2021saturation}, and allows us to compare our findings with previous studies. The turbulent acceleration field is constructed in Fourier space, which allows us to isotropically inject energy into wavenumbers, $k = |\bm{k}|$, at an effective driving scale of the turbulence on the box scale, $\ell_\turb = L$, which corresponds to $k_\turb = 2\pi/\ell_\turb$. Throughout this study, we will report wavenumbers in units of $k_\turb$. We drive over wavenumbers $0.5 < k < 1.5$ with a parabolic spectrum for the Fourier amplitudes, which peaks at $k = 1$, and is zero at $k = 0.5$ and $k = 1.5$.

The auto-correlation time of $\bm{F}$ is $\teddy = \ell_\turb / (\Mach c_s)$, and the driving amplitude is adjusted so that the desired sonic Mach number, $\Mach = u_\turb / c_s \approx 0.3$ is achieved in the kinematic phase of the dynamo for all of our simulations (see \autoref{tab:simulations_setup}). As our simulations are for subsonic turbulence, density fluctuations in all of our simulations are relatively small, with $\delta\rho / \rho_0$ of the order of 0.1--1\%. Thus, $\rho$ can be considered approximately constant, $\rho\approx\rho_0$.

\subsection{Initial conditions and plasma Reynolds numbers} \label{subsec:initalise}

\begin{table*}
    \renewcommand{\arraystretch}{1.3}
    \setlength{\tabcolsep}{2.1pt}
    \caption{Main simulation parameters and derived quantities.}
    \label{tab:simulations_setup}
    \begin{tabular}{l l l l c c c c c c c c c c}
        \hline\hline
                Simulation ID & \Reyk & \Reym & \Pranm
                & $\nu$ & $\eta$
                & $\Mach$ & $\grate$ & $\rbrac{E_\tmag / E_\tkin}_\text{sat}$ &
                $\alpha_\tkin$ & $\alpha_\tmag$ & $\knu$ & $\keta$ & $\kp$ \\
                (1) & (2) & (3) & (4) & (5) & (6) & (7) & (8) & (9) & (10) & (11) & (12) & (13) & (14) \\
            \hline\hline
            \multicolumn{14}{c}{$\Reyk = 10$} \\
            \hline
            Re10Pm27$^\dag$
                & $10$ & $270$ & $27$
                & $2.50\times 10^{-2}$ & $1.00\times 10^{-3}$
                & $0.27_{-0.03}^{+0.05}$ & decaying & --
                & $-3_{-2}^{+2}$ & $3.0_{-0.2}^{+0.2}$
                & $2.0_{-0.2}^{+0.2}$ & $2.8_{-0.2}^{+0.6}$ & $3.0_{-0.1}^{+0.1}$ \\
            Re10Pm54$^\dag$
                & $10$ & $540$ & $54$
                & $2.50\times 10^{-2}$ & $5.00\times 10^{-4}$
                & $0.27_{-0.03}^{+0.05}$ & $0.29 \pm 0.03$ & $0.17_{-0.07}^{+0.33}$
                & $-3_{-2}^{+2}$ & $2.9_{-0.3}^{+0.3}$
                & $0.30_{-0.01}^{+0.02}$ & $1.1_{-0.1}^{+0.1}$ & $2.7_{-0.3}^{+0.3}$ \\
            Re10Pm130$^\dag$
                & $10$ & $1300$ & $130$
                & $2.50\times 10^{-2}$ & $2.00\times 10^{-4}$
                & $0.26_{-0.03}^{+0.06}$ & $0.61 \pm 0.04$ & $1.1_{-0.3}^{+0.6}$
                & $-3_{-2}^{+2}$ & $2.8_{-0.4}^{+0.4}$
                & $0.28_{-0.02}^{+0.02}$ & $1.8_{-0.4}^{+0.4}$ & $4.0_{-0.8}^{+0.8}$ \\
            Re10Pm250$^\dag$
                & $10$ & $2500$ & $250$
                & $2.50\times 10^{-2}$ & $1.00\times 10^{-4}$
                & $0.25_{-0.05}^{+0.06}$ & $0.66 \pm 0.02$ & $1.9_{-0.5}^{+1.0}$
                & $-3_{-2}^{+2}$ & $2.7_{-0.4}^{+0.4}$
                & $0.30_{-0.01}^{+0.01}$ & $2.2_{-0.5}^{+0.6}$ & $5_{-1}^{+1}$ \\
            \hline
            \multicolumn{14}{c}{$\Reyk \approx 450$} \\
            \hline
            Re430Pm1
                & $430$ & $430$ & $1$
                & $6.00\times 10^{-4}$ & $6.00\times 10^{-4}$
                & $0.26_{-0.02}^{+0.07}$ & decaying & --
                & $-1_{-2}^{+3}$ & $1.6_{-0.4}^{+0.4}$
                & $2.2_{-0.2}^{+0.2}$ & $2.2_{-0.7}^{+0.7}$ & $2.4_{-0.2}^{+0.2}$ \\
            Re470Pm2$^\dag$
                & $470$ & $940$ & $2$
                & $6.00\times 10^{-4}$ & $3.00\times 10^{-4}$
                & $0.28_{-0.03}^{+0.04}$ & $0.37 \pm 0.02$ & $0.11_{-0.04}^{+0.06}$
                & $-1_{-2}^{+3}$ & $1.7_{-0.4}^{+0.4}$
                & $2.2_{-0.1}^{+0.1}$ & $2.9_{-0.3}^{+0.2}$ & $3.4_{-0.3}^{+0.3}$ \\
            Re470Pm4
                & $470$ & $1900$ & $4$
                & $6.00\times 10^{-4}$ & $1.50\times 10^{-4}$
                & $0.28_{-0.06}^{+0.03}$ & $0.69 \pm 0.04$ & $0.2_{-0.1}^{+0.1}$
                & $-1_{-3}^{+3}$ & $1.7_{-0.3}^{+0.3}$
                & $2.2_{-0.2}^{+0.2}$ & $4.0_{-0.8}^{+0.8}$ & $4.4_{-0.4}^{+0.4}$ \\
            \hline
            \multicolumn{14}{c}{$\Reym \approx 3300$} \\
            \hline
            Re3600Pm1
                & $3600$ & $3600$ & $1$
                & $8.33\times 10^{-5}$ & $8.33\times 10^{-5}$
                & $0.30_{-0.03}^{+0.03}$ & $1.1 \pm 0.1$ & $0.12_{-0.02}^{+0.02}$
                & $-1_{-1}^{+1}$ & $1.9_{-0.2}^{+0.2}$
                & $11_{-4}^{+4}$ & $8_{-2}^{+2}$ & $13_{-3}^{+3}$ \\
            Re1700Pm2$^\dag$
                & $1700$ & $3400$ & $2$
                & $1.67\times 10^{-4}$ & $8.33\times 10^{-5}$
                & $0.31_{-0.05}^{+0.04}$ & $1.3 \pm 0.1$ & $0.21_{-0.05}^{+0.09}$
                & $-1_{-1}^{+1}$ & $1.7_{-0.2}^{+0.2}$
                & $6.6_{-0.5}^{+0.4}$ & $8_{-1}^{+1}$ & $9.6_{-0.8}^{+0.8}$ \\
            Re600Pm5
                & $600$ & $3000$ & $5$
                & $4.17\times 10^{-4}$ & $8.33\times 10^{-5}$
                & $0.25_{-0.03}^{+0.04}$ & $1.5 \pm 0.1$ & $0.23_{-0.05}^{+0.09}$
                & $-1_{-2}^{+3}$ & $1.8_{-0.3}^{+0.3}$
                & $3.2_{-0.3}^{+0.3}$ & $6_{-1}^{+2}$ & $7_{-2}^{+2}$ \\
            Re290Pm10$^\dag$
                & $290$ & $2900$ & $10$
                & $8.33\times 10^{-4}$ & $8.33\times 10^{-5}$
                & $0.24_{-0.03}^{+0.06}$ & $1.3 \pm 0.1$ & $0.4_{-0.1}^{+0.2}$
                & $-1_{-2}^{+2}$ & $1.7_{-0.2}^{+0.2}$
                & $1.9_{-0.1}^{+0.1}$ & $5.3_{-0.6}^{+0.6}$ & $5.8_{-0.6}^{+0.6}$ \\
            Re140Pm25
                & $140$ & $3500$ & $25$
                & $2.08\times 10^{-3}$ & $8.33\times 10^{-5}$
                & $0.29_{-0.05}^{+0.06}$ & $2.3 \pm 0.1$ & $0.5_{-0.1}^{+0.2}$
                & $-2_{-1}^{+1}$ & $1.8_{-0.3}^{+0.3}$
                & $1.3_{-0.2}^{+0.2}$ & $5_{-1}^{+2}$ & $6_{-1}^{+1}$ \\
            Re64Pm50
                & $64$ & $3200$ & $50$
                & $4.17\times 10^{-3}$ & $8.33\times 10^{-5}$
                & $0.27_{-0.05}^{+0.04}$ & $0.93 \pm 0.05$ & $0.9_{-0.3}^{+0.4}$
                & $-1_{-1}^{+2}$ & $2.1_{-0.2}^{+0.2}$
                & $0.61_{-0.04}^{+0.04}$ & $4_{-1}^{+1}$ & $5.0_{-0.8}^{+0.7}$ \\
            Re27Pm130$^\dag$
                & $27$ & $3500$ & $130$
                & $1.04\times 10^{-2}$ & $8.33\times 10^{-5}$
                & $0.29_{-0.06}^{+0.04}$ & $0.98 \pm 0.06$ & $1.4_{-0.5}^{+0.6}$
                & $-1_{-1}^{+1}$ & $2.4_{-0.3}^{+0.3}$
                & $0.36_{-0.02}^{+0.02}$ & $3.1_{-0.6}^{+0.6}$ & $5.6_{-0.7}^{+0.6}$ \\
            Re12Pm260$^\dag$
                & $12$ & $3100$ & $260$
                & $2.08\times 10^{-2}$ & $8.33\times 10^{-5}$
                & $0.26_{-0.06}^{+0.04}$ & $0.89 \pm 0.06$ & $2.0_{-0.6}^{+1.7}$
                & $-2_{-2}^{+2}$ & $2.6_{-0.3}^{+0.3}$
                & $0.32_{-0.04}^{+0.04}$ & $3.2_{-0.5}^{+0.5}$ & $5.9_{-0.4}^{+0.4}$ \\
            \hline
            \multicolumn{14}{c}{$\ketatheory \approx 125$} \\
            \hline
            Re73Pm26
                & $73$ & $1900$ & $26$
                & $3.38\times 10^{-3}$ & $1.35\times 10^{-4}$
                & $0.25_{-0.03}^{+0.05}$ & $0.61 \pm 0.03$ & $0.4_{-0.1}^{+0.2}$
                & $-1_{-1}^{+3}$ & $2.1_{-0.3}^{+0.3}$
                & $0.67_{-0.06}^{+0.06}$ & $2.5_{-0.5}^{+0.6}$ & $3.9_{-0.3}^{+0.3}$ \\
            Re48Pm52
                & $48$ & $2500$ & $52$
                & $5.32\times 10^{-3}$ & $1.06\times 10^{-4}$
                & $0.26_{-0.04}^{+0.04}$ & $0.81 \pm 0.04$ & $0.7_{-0.3}^{+0.4}$
                & $-1_{-1}^{+1}$ & $2.2_{-0.3}^{+0.3}$
                & $0.52_{-0.04}^{+0.04}$ & $2.7_{-0.8}^{+0.8}$ & $4.2_{-0.5}^{+0.6}$ \\
            Re25Pm140
                & $25$ & $3500$ & $140$
                & $9.74\times 10^{-3}$ & $7.79\times 10^{-5}$
                & $0.27_{-0.06}^{+0.03}$ & $0.86 \pm 0.04$ & $1.3_{-0.4}^{+0.6}$
                & $-1_{-1}^{+1}$ & $2.3_{-0.3}^{+0.3}$
                & $0.38_{-0.03}^{+0.03}$ & $3_{-1}^{+1}$ & $5_{-2}^{+1}$ \\
            Re16Pm250$^\dag$
                & $16$ & $4000$ & $250$
                & $1.56\times 10^{-2}$ & $6.25\times 10^{-5}$
                & $0.25_{-0.07}^{+0.04}$ & $1.0 \pm 0.1$ & $1.9_{-0.7}^{+1.2}$
                & $-1_{-1}^{+1}$ & $2.5_{-0.3}^{+0.3}$
                & $0.31_{-0.04}^{+0.04}$ & $3.4_{-0.8}^{+0.7}$ & $6_{-2}^{+1}$ \\
            \hline\hline
    \end{tabular}
    \begin{tablenotes}[para]
        \textit{Note:} All derived quantities (with the exception of the saturated energy) are time averaged over a subset of time realisations within the kinematic phase of the dynamo, namely, where $10^{-6} \leq E_\tmag/E_\tkin \leq 10^{-2}$. Columns:
        \textbf{(1)}: The simulation ID, where $^\dag$ indicates those simulations that have been run at a resolution of $\Nres = 576$ in addition to the default resolutions of $\Nres = 18$, $36$, $72$, $144$ and $288$ (which are common to all runs).
        \textbf{(2)}: The hydrodynamic Reynolds number (see \autoref{eqn:Reyk}).
        \textbf{(3)}: The magnetic Reynolds number (see \autoref{eqn:Reym}).
        \textbf{(4)}: The magnetic Prandtl number (see \autoref{eqn:Pranm}).
        \textbf{(5)}: The kinematic viscosity in units of $\ell_\turb^2 / \teddy$.
        \textbf{(6)}: The magnetic resistivity in units of $\ell_\turb^2 / \teddy$.
        \textbf{(7)}: The measured turbulent velocity (Mach number) during the kinematic phase.
        \textbf{(8)}: The measured growth rate, in units of $\teddy^{-1}$, of the magnetic energy during the kinematic phase.
        \textbf{(9)}: The measured ratio between the magnetic and kinetic energy in the saturated stage of the dynamo.
        In the next five columns we report the velocity and magnetic spectra power-law exponents, as well as characteristic wavenumbers measured directly from spectra (see \S\ref{subsec:spectra_fits}). The wavenumbers are reported in units of $k_\turb$ (see \S\ref{subsec:turb_driving}).
        \textbf{(10)}: The measured power-law exponent for the kinetic spectra.
        \textbf{(11)}: The measured power-law exponent for the magnetic spectra.
        \textbf{(12)}: The measured viscous dissipation wavenumber.
        \textbf{(13)}: The measured resistive dissipation wavenumber.
        \textbf{(14)}: The wavenumber associated with the peak magnetic energy.
    \end{tablenotes}
\end{table*}

We initialise all our simulations with constant density, zero velocity, $\bm{u} = \bm{0}$, and zero mean magnetic field, $\bm{B}_0 = \bm{0}$, so $\bm{B} = \delta\bm{B}$. We initialise $\bm{B}$ over the largest scales in the simulation domain, $0.5 < k < 1.5$, with a parabolic profile that peaks at $k = 1$, and is zero at $k = 0.5$ and $k = 1.5$ (which is the same profile that we drive turbulence with; see the previous section). For all our simulations, we choose the initial field such that the plasma $\beta \equiv p_\text{th}/p_\tmag = 10^{10}$, where $p_\tmag = B^2/(8\pi)$. \citet{seta2020seed} showed that all properties of dynamo-generated magnetic fields are not affected by the initial seed field structure or strength (as long as the field is weak).

We study simulations in the \mbox{$\Pranm \geq 1$} regime with \mbox{$\Reyk=10$--$3600$}. All our simulations are evolved until \mbox{$t/\teddy = 100$}, well into the saturated regime of the dynamo. We test numerical convergence by using different linear grid resolutions, $\Nres = 18$, $36$, $72$, $144$, $288$, and $576$.

We run four different sets of simulations, grouped in \autoref{tab:simulations_setup}. First, we run four simulations similar to \citet{schekochihin2004simulations}\footnote{
    \citet{schekochihin2004simulations} defined the hydrodynamic Reynolds number with respect to the driving wavenumber, and thus, the $\Reyk$ that they report is lower than the ones we do by a factor of $2\pi$.
}, where $\Reyk = 10$ is fixed for all simulations, and $\Reym$ is varied in order to achieve \mbox{$\Pranm \approx 25$--$250$}. Second, we run three simulations, where we fix $\Reyk \approx 450$, and vary $\Reym$ to achieve $\Pranm = 1$, $2$, and $4$. Third, we run eight simulations where we fix \mbox{$\Reym \approx 3300$}, and vary $\Reyk$ to achieve $\Pranm = 1$--$260$. Finally, to test whether the dependence of $\kp$ is on $\ketatheory$, we run four simulations where we fix $\ketatheory \approx 125$, and vary $\Reyk$ and $\Reym$ to achieve \mbox{$\Pranm \approx 25$--$250$}. We report all relevant simulation parameters and derived quantities in \autoref{tab:simulations_setup}. 

Throughout this study, we use dimensionless units to describe physical quantities: $\rho$ is in units of $\rho_0$, $\bm{u}$ is in units of $c_s$, and $\bm{B}$ is in units of $c_s\rho_0^{1/2}$. For all simulations, we set $c_s = \rho_0 = 1$. Our dissipation coefficients, $\nu$ and $\eta$, are reported in units of $\ell_\turb^2 / \teddy$.

%% RESULTS
\section{Results} \label{sec:results}

\subsection{Time evolution and basic properties of the turbulent dynamo} \label{subsec:B_evolution}

We start by comparing two representative simulation models (Re470Pm2 and Re1700Pm2), in order to highlight some of the fundamental differences in the properties of amplified magnetic fields in low- and high-$\Reyk$ turbulent flows, and to introduce the analysis methods that we ultimately apply to all of our simulations. However, before going into the details about how we measure important length scales of the turbulent dynamo, we first confirm that magnetic field amplifies in our simulations.

\begin{figure}
    \centering
    \includegraphics[width=\linewidth]{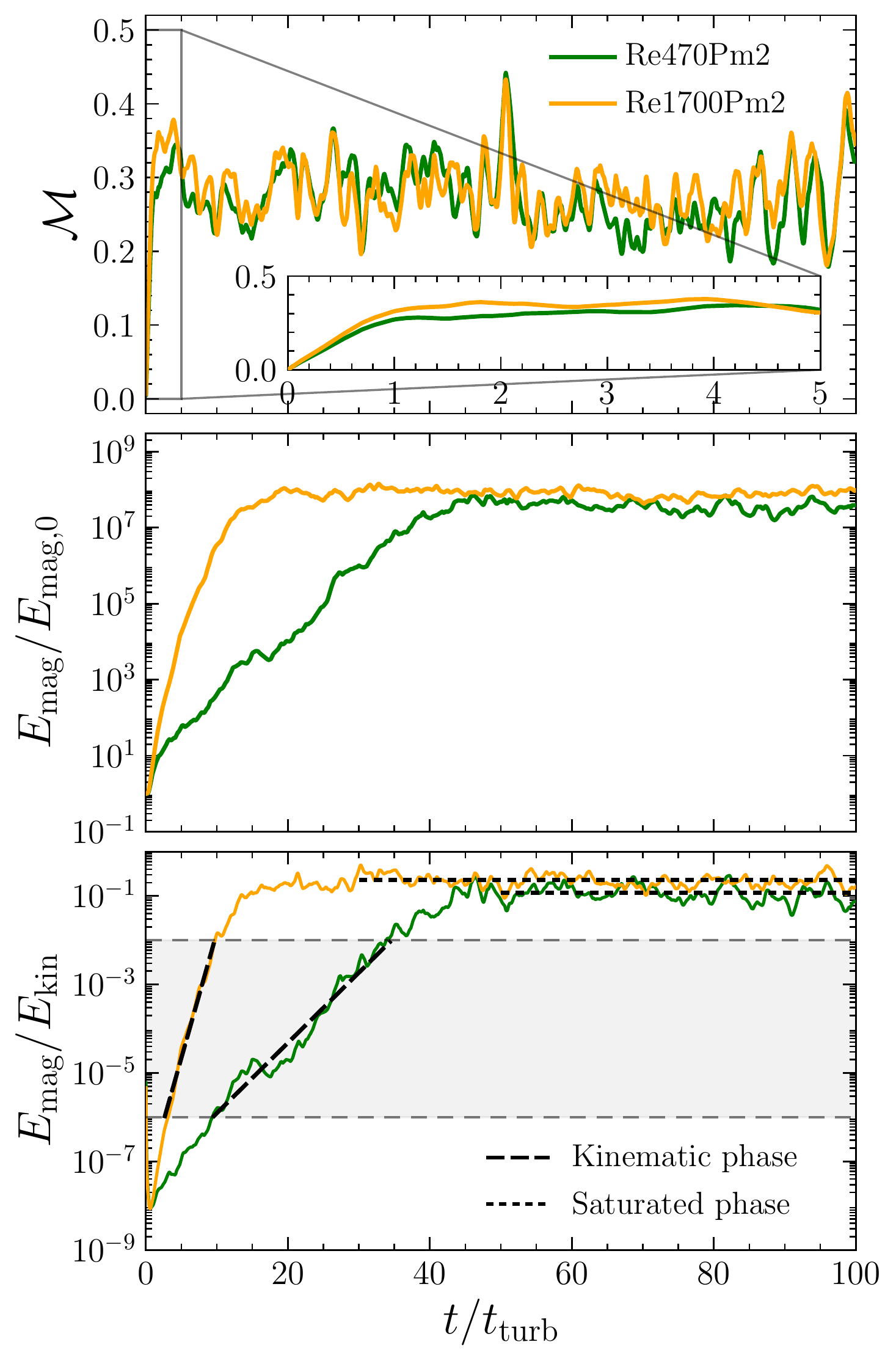}
    \caption{Time evolution of the sonic Mach number, $\Mach$ (top panel), magnetic energy normalised by the initial magnetic energy, $E_\tmag / E_{\tmag, 0}$ (middle panel), and the ratio of magnetic to kinetic energy, $E_\tmag / E_\tkin$ (bottom panel), for our Re470Pm2 (green) and Rm1700Pm2 (yellow) simulations. After an initial transient phase (see inset in the top panel), $t/\teddy \approx 2$, the turbulence in the simulation domain is fully developed. In the kinematic phase of the dynamo, we fit an exponential function, $\exp(\grate\, t/\teddy)$ (dashed line), to the evolution of the energy ratio (bottom panel) over time realisations where $10^{-6} \leq E_\tmag / E_\tkin \leq 10^{-2}$ (indicated by the grey band). In the saturated phase of the dynamo, we measure the saturation level, $\satlevel$ (dotted line). We report the growth rate, $\grate$, and the saturation level we measured for all of our simulations in \autoref{tab:simulations_setup}.}
    \label{fig:energy_evol}
\end{figure}

In \autoref{fig:energy_evol}, we show the time evolution of the sonic Mach number, $\Mach$ (top panel), the magnetic energy, normalised by its initial value, $E_\text{mag} / E_{\text{mag}, 0}$ (middle panel), and the ratio of magnetic to kinetic energy, $E_\text{mag} / E_\text{kin}$ (bottom panel) for the Re470Pm2 and Re1700Pm2 simulations. After an initial transient period, $t/\teddy \approx 2$, the turbulence is fully developed (see inset in the top panel of \autoref{fig:energy_evol}). We note that for MHD turbulence with a strong mean field, this transient phase can take up to $t \approx 5\, \teddy$ to become fully developed \citep{beattie2021_spdf}, while for hydrodynamical supersonic turbulence, this time is somewhat shorter, $t \approx 2\, \teddy$ \citep{federrath2010comparing, price2010comparison}. We make sure that all of our statistics are calculated from time realisations after this transient phase. We measure that $\Mach$ becomes statistically stationary at a value of $\Mach\approx0.3$ for the Re470Pm2 and Re1700Pm2 simulations. We measure $\Mach$ for all of our simulations (see column (7) in \autoref{tab:simulations_setup}) and find that they all lie within $\approx 10$\% of our target $\Mach = 0.3$.

The middle panel of \autoref{fig:energy_evol} shows the evolution of the magnetic energy. We see that the initially weak seed magnetic field grows exponentially (kinematic phase), achieving more than six orders of magnitude of magnetic amplification for both of the example simulations. However, this is true for all of our simulations where we measure magnetic amplification. Throughout this study, we calculate statistics in the kinematic phase of the dynamo by averaging over all time realisations where $10^{-6} \leq E_\tmag/E_\tkin \leq 10^{-2}$ (indicated by the shaded grey band in the bottom panel of \autoref{fig:energy_evol}). This averaging range lies within the kinematic phase for all our amplifying simulations and ensures that we measure the growth rates and fundamental length scales sufficiently far away from the initial transient phase and the saturated phase.

We find that for all our simulations where the scale separation between $\knu$ and $\keta$ is fixed, an increase in $\Reyk$ corresponds to faster amplification of the magnetic field. This result aligns with theoretical expectations, that in the $\Pranm\gg 1$ limit, for \citet{Kolmogorov1941} turbulence ($u_\turb\propto\ell^{1/3}$), the growth rate scales like $\grate\propto\Reyk^{1/2}$ \citep{batchelor1950spontaneous, kulsrud1992spectrum, haugen2004simulations, schekochihin2004simulations, schober2012magnetic}. \citet{bovino2013turbulent} formulated a semi-analytic model for $\grate$ as a function of the velocity scaling exponent, $\vartheta$, $\Reyk$, and $\Pranm$. We evaluate their model for \citet{Kolmogorov1941} turbulence, with $\vartheta=1/3$, and find $\grate \approx 0.39$ for $\Reyk=470$ and $\grate \approx 0.98$ for $\Reyk=1700$, with $\Pranm=2$ in both cases. From \autoref{fig:energy_evol}, we measure $\grate = 0.37 \pm 0.02$ and $1.3 \pm 0.1$ for the Re470Pm2 and Re1700Pm2 simulations, respectively. Considering the overall factor of 3--4 difference in the growth rates of these two simulations, the agreement with the theoretical predictions is very good. The small differences ($\lesssim30\%$) between the measured and predicted $\grate$ could be the result of the implicit assumption of delta-correlated driving in the \citet{bovino2013turbulent} model, whereas our simulations have finite-correlated driving \citep[see also discussion in][]{lim2020generation}.

Once the magnetic field is strong enough to suppress the turbulent stretching motions \citep{schekochihin2004simulations, SetaEA2020sat, seta2021saturation}, magnetic amplification slows down and reaches a final saturated state (saturated phase). We measure a statistically saturated level of $\satlevel = 0.11_{-0.04}^{+0.06}$ for the Re470Pm2 simulation, and $0.21_{-0.05}^{+0.09}$ for Re1700Pm2. Thus, for these two simulations, the magnetic energy reaches $\approx10$--$20$\% of the turbulent kinetic energy. For solenoidally driven turbulence, \citet{federrath2011mach} provides an empirical model that predicts $\satlevel=0.44$ for $\Mach=0.3$, $\Reyk=1500$, and $\Pranm=2$. When we compare this with our Re1700Pm2 simulation, we find that we measure $\satlevel$ that is a factor of $\approx2$ lower. This difference is likely a consequence of our simulations being driven on $k_\turb = 1$ (i.e., $\ell_\turb = L$), whereas \citet{federrath2011mach} drive their turbulence on $k_\turb = 2$ (i.e., $\ell_\turb = L/2$).

While there is currently no analytical model for the saturation level that predicts $\satlevel$ as a function of the plasma Reynolds numbers, current simulations suggest that the saturation level depends on $\Pranm$ and $\Reyk$ \citep{schekochihin2004simulations, schober2015saturation}\footnote{
    Note that theoretically the dependence of the saturation level upon $\Pranm$ and $\Reyk$ is a repercussion of the finite plasma parameters, and need not hold in the $\Pranm\rightarrow0$ and $\Pranm\rightarrow\infty$ limits.
}, even for supersonic turbulence \citep{federrath2014turbulent}. Finally, we also find that our set of simulations where $\Reyk=10$ and $\Pranm=54$--$250$ (see the first four runs presented in \autoref{tab:simulations_setup}), agrees with the saturation levels that \citet{schekochihin2004simulations} report for similar simulation parameters. We measure $\grate$ and $\satlevel$ for all of our simulations, and report them in columns (8) and (9) in \autoref{tab:simulations_setup}, respectively.

\subsection{Kinetic and magnetic energy structures} \label{subsec:slices}

\begin{figure}
    \centering
    \includegraphics[width=\linewidth]{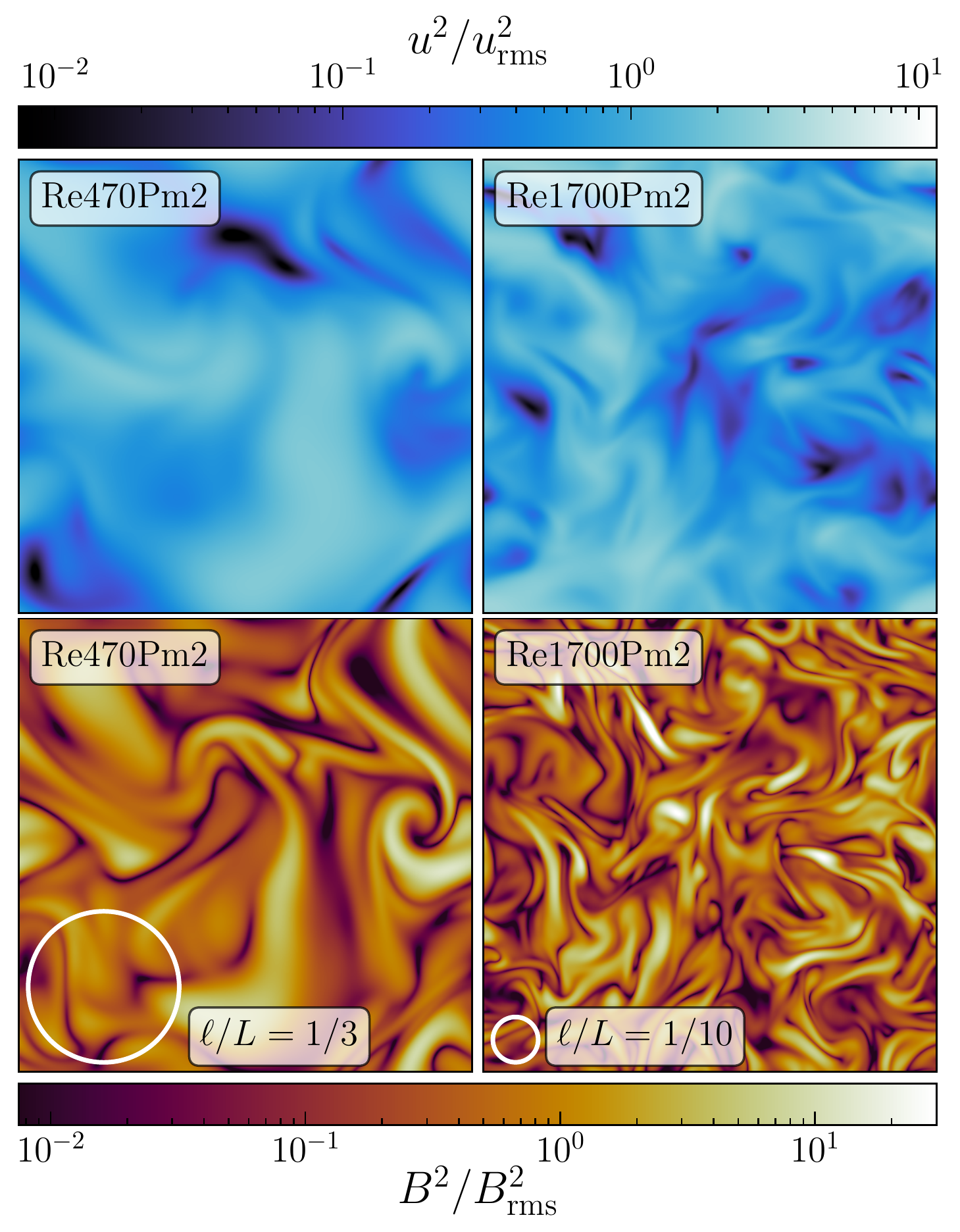}
    \caption{
        Two-dimensional slices taken of the $u^2/u_\rms^2$ (top panels) and $B^2/B_\rms^2$ (bottom panels) fields, for our Re470Pm2 and Re1700Pm2 simulations, respectively, where $u_\rms$ and $B_\rms$ are the root-mean-square of the velocity and magnetic fields, respectively. These slices are from the middle of our simulation box domain, $(x, y, z=L/2)$, at the time realisation where the energy ratio $E_\text{mag} / E_\text{kin} = 10^{-4}$. These slices have been normalised by the root-mean-squared (rms) values to reveal the morphological structures in each of the fields.
    }
    \label{fig:energy_slice}
\end{figure}

In \autoref{fig:energy_slice} we show two-dimensional slices of $u^2/u_\rms^2$ (top panels), where $u_\rms$ is the root-mean-squared (rms) velocity, and $B^2 / B_\rms^2$ (bottom panels), where $B_\rms$ is the rms magnetic field, for our Re470Pm2 (left-hand panels) and Re1700Pm2 (right-hand panels) simulations. Since $\ave{\bm{u}} = 0$ in our simulations, $u_\rms = u_\turb$. We normalise by the rms velocity and magnetic field, respectively, because we are only interested in the structure of the fields, rather than the magnitude of the fields. All four slices are taken from the middle of the box domain, $(x, y, z=L/2)$, at the time realisation where $E_\tmag / E_\tkin = 10^{-4}$, which corresponds to the kinematic phase of the dynamo. As discussed previously (see \S\ref{subsec:turb_driving}), the density fluctuations are small, $\rho \approx \rho_0$, in our simulations, and therefore $u^2/u_\rms^2$ is proportional to $(1/2)~\rho_0 u^2$, and hence the square of the velocity structures are equivalent to the kinetic energy structures. For the remainder of this section, we will refer to $u^2/2$ as the turbulent kinetic energy.

Visually, both the turbulent kinetic and magnetic energy densities appear to be concentrated on smaller scales in the Re1700Pm2 simulation, compared with the Re470Pm2 simulation, where $\Reyk = 1700$ and $470$, respectively, with $\Pranm = 2$. It is well understood that when $\Reyk$ increases, then the viscous scale eddies (theoretically given by \autoref{eqn:knutheory}) shift to smaller scales, which increases the range of scales that the scale-free energy cascade spans \citep{Kolmogorov1941}. For both of these simulations where $\Pranm=2$ has been fixed, we also see that magnetic field energy densities are more concentrated on smaller scales for the higher-Re simulation. We have indicated in \autoref{fig:energy_slice} circles with radius equal to the length scale where it appears that the magnetic field energies are predominantly concentrated. Specifically, For Re470Pm2, magnetic energy appears concentrated roughly at a third of the box length, $\ell/L = 1/3$, and for Re1700Pm2 at around a tenth of the box length, $\ell/L = 10$ (in \S\ref{subsec:spectra_fits} we quantify these scales by measuring $\kp$ in our simulations).

For both the representative simulations, the magnetic field energy density appears to be concentrated at scales smaller than the turbulent kinetic energy density. In the next section, we discuss our method for measuring fundamental length scales in both the kinetic and magnetic energy fields, including the peak scale of the magnetic field.

\subsection{Kinetic and magnetic power spectra} \label{subsec:spectra_fits}

To determine on which scale magnetic fields become most concentrated, we study the functional form of the velocity and magnetic power spectra, and characterise the spectra by measuring $\knu$, $\keta$ and $\kp$. This allows us to determine whether $\kp$ depends on $\knu$ \citep{batchelor1950spontaneous} or on $\keta$ \citep{kazantsev1968enhancement, kulsrud1992spectrum, Vainshtein1992, schekochihin2002spectra, schekochihin2004simulations, brandenburg2005astrophysical, schober2015saturation, xu2016turbulent, mckee2020magnetic}.

In the kinematic phase of the dynamo, the kinetic energy is significantly greater than the magnetic energy, and the kinetic energy spectrum is largely unaffected by the magnetic spectra. In our simulations (as discussed in \S\ref{subsec:turb_driving}), fluctuations in the density field are small, and therefore the velocity power spectra are proportional to the kinetic energy spectra. In the remainder of this study, we will refer to the velocity power spectra as the kinetic energy spectra. We propose a simple model for the kinetic spectra, which is motivated by the shape of the spectrum in the kinematic phase. For \citet{Kolmogorov1941} turbulence, the kinetic energy spectrum consists of a power law, which spans over the inertial range $k_\turb \ll k \ll \knu$. Beyond $\knu$, dissipation dominates, which we model with a decaying exponential function. Our model for the kinetic energy spectrum is
\begin{equation}
    \Pvel(k) = A_\tkin \;k^{\alpha_\tkin} \exp\rbrac{ -k / \knu } , \label{eqn:kin_spectra}
\end{equation}
where $A_\tkin$ is a constant, $\alpha_\tkin$ is the slope of the power law in the scaling range, and $\knu$ is the dissipation wavenumber. Note that the expectation is $\alpha_\tkin = -5/3$ (ignoring intermittency effects, e.g. \citealt{She1994_intermittency}) for \citet{Kolmogorov1941} turbulence, but here it is a free parameter to be determined from fits of this model to the velocity spectra of our simulations. However, in \aref{app:kolmog_exp} we also test the effects of fixing $\alpha_\tkin = -5/3$ and find that it does not significantly affect our measurements of $\knu$, which is the main fit parameter in this model, for the purposes of this study.

To model the magnetic power spectra, we use a solution to the Kazantsev equation \citep{kazantsev1968enhancement, brandenburg2005astrophysical} for the kinematic phase of the dynamo, as derived by \citet{kulsrud1992spectrum}. The Kazantsev model assumes an isotropic, homogeneous, Gaussian random velocity field, with zero helicity, and $\delta$-correlation in time. The functional form of the magnetic power spectrum is given as \citep{kulsrud1992spectrum}, 
\begin{equation}
    \Pmag(k) = A_\tmag \;k^{\alpha_\tmag} K_0\rbrac{ k / \keta } , \label{eqn:mag_spectra}
\end{equation}
where $A_\tmag$ is a constant, $\alpha_\tmag$ is the slope of the power law, and $K_0$ is the modified Bessel function of the second kind and order $0$. The slope of the power law in the solution to the Kazantsev equation is $3/2$, but like the kinetic energy model, we retain it as a free parameter to explicitly measure the exponent in our simulations.

For all of our simulations, we fit the kinetic and magnetic spectra with \autoref{eqn:kin_spectra} and \ref{eqn:mag_spectra}, respectively, to each time realisation where $10^{-6} \leq E_\text{mag} / E_\text{kin} \leq 10^{-2}$, corresponding to the kinematic phase of the dynamo. For each of these fits, we measure the dissipation wavenumbers, $\knu$ and $\keta$, from the fitted spectra. We also measure $\kp$, the peak of the magnetic power spectra, analytically by finding where the first derivative of the magnetic spectra is zero, 
\begin{equation}
    \driv{\Pmag}{k}
    \equiv A_\tmag k^{\alpha_\tmag} \sbrac{
        \frac{\alpha_\tmag}{k} K_0\rbrac{\frac{k}{\keta}} - \frac{1}{\keta} K_1\rbrac{\frac{k}{\keta}} 
    }
    = 0 , 
\end{equation}
where $K_1(x)$ is the modified Bessel function of the second kind and order $1$. With the requirement that $A_\tmag \ne 0$, the only nontrivial relation for $\kp$ that follows from this is, 
\begin{equation}
    \kp = \alpha_\tmag\, \frac{K_0(\kp / \keta)}{K_1(\kp / \keta)} \keta . \label{eqn:kp_implicit}
\end{equation}
This equation implicitly relates $\kp$ and $\keta$ via a constant of proportionality that involves $\alpha_\tmag$ and the fraction of two modified Bessel functions. Since, $0 < K_0(x)/K_1(x) < 1$ for all $x > 0$, the constant of proportionality is bounded between $0$ and $\alpha_\tmag$.

\begin{figure}
    \centering
    \includegraphics[width=\linewidth]{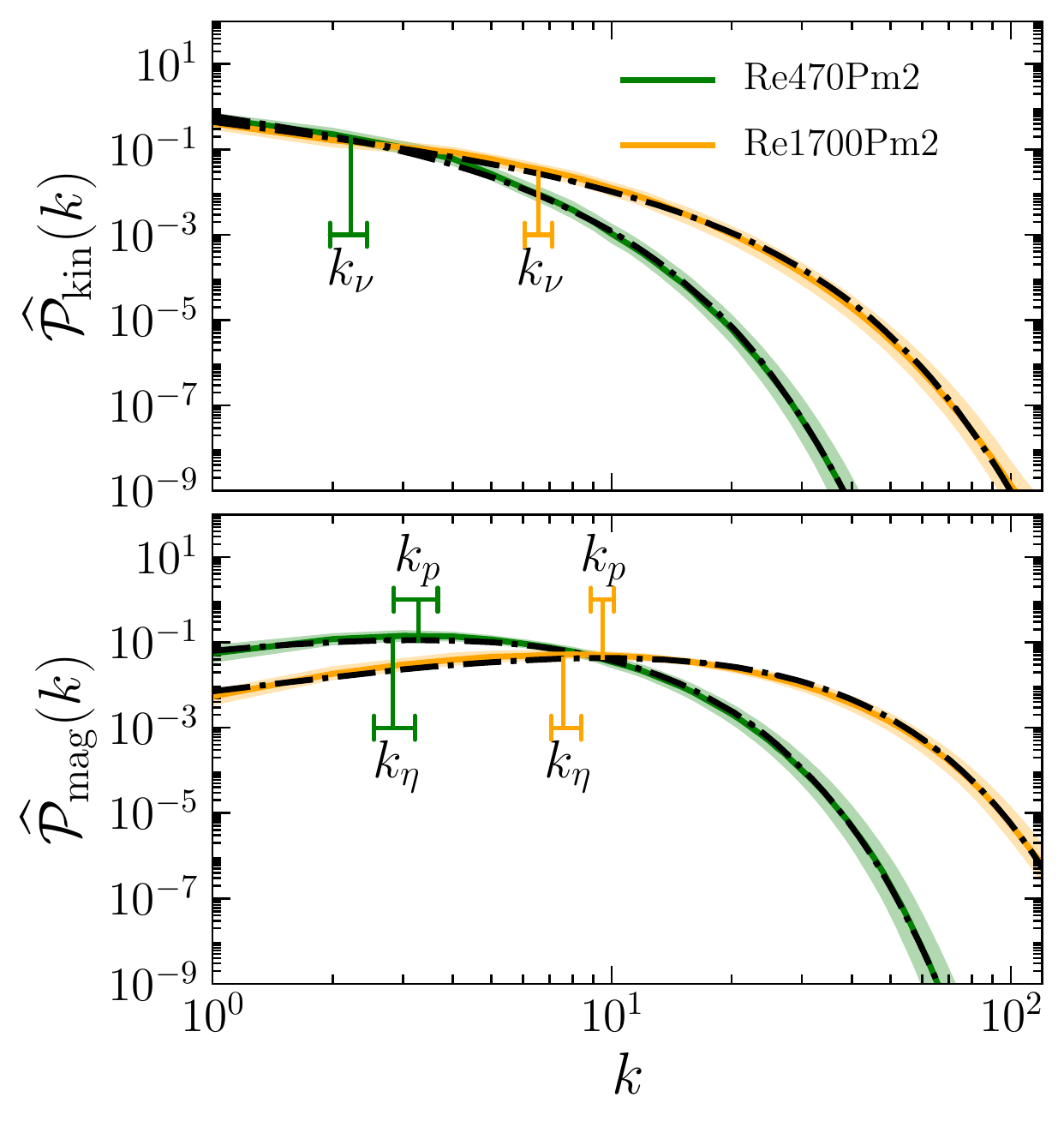}
    \caption{The normalised and time-averaged kinetic energy spectra, $\mathcal{\widehat{P}}_\tkin(k)$ (top panel), and magnetic power spectra, $\mathcal{\widehat{P}}_\tkin(k)$ (bottom panel), for our Re470Pm2 and Re1700Pm2 simulations. We overlay our fitted spectra models (see \autoref{eqn:kin_spectra} and \autoref{eqn:mag_spectra}) shown with a black dash-dotted line, and annotate the measured dissipation wavenumbers, $\knu$ and $\keta$, as well as the peak magnetic energy scale, $\kp$. See columns (10) and (11) in \autoref{tab:simulations_setup} for the time-averaged $\alpha_\tkin$ and $\alpha_\tmag$ reported for each simulation, respectively, and columns (12 -- 14) for the time averaged $\knu$, $\keta$ and $\kp$, respectively. We show the uncertainty of the measured scales with a bracket that spans between the $16$th and $84$th percentiles.}
    \label{fig:spectra}
\end{figure}

In \autoref{fig:spectra}, we show these spectra models fitted to the time-averaged and normalised kinetic energy spectra (top panel) and magnetic power spectra (bottom panel), for the Re470Pm2 (green) and Re1700Pm2 (yellow) simulations. We overlay our spectra fits with a black dash-dotted line, and annotate the measured $\knu$, $\keta$ and $\kp$, indicating the uncertainty in these scales with the width of the bracket.

As indicated in the bottom panel of \autoref{fig:spectra}, we measure $\keta$ on smaller wavenumbers (larger scales) than $\kp$ for both the Re470Pm2 and Re1700Pm2 simulations. This is also true for all of our simulations (see columns (13) and (14) in \autoref{tab:simulations_setup}). We emphasise that $\knu$ and $\keta$ are characteristic wavenumbers, where the dissipation terms in our spectral models ($\Pvel$ and $\Pmag$, respectively) start to dominate. However, it does not mean, for example, that $\Pmag(k \geq k_\eta) = 0$. In fact, the magnetic spectrum is typically peaked at the resistive scale, $\kp \gtrsim \keta$, as we will see later. Therefore, these characteristic wavenumbers are what we measure as our dissipation wavenumbers.

To ensure that the estimated scales are numerically converged with respect to the resolution of the simulation, we perform a scale convergence analysis in the next section.

\subsection{Scale convergence} \label{subsec:resolution}

Before we study the dependence of $\kp$ on $\knu$ and $\keta$, we ensure that we work with scales that have numerically converged. In this section, we present our resolution study of $\knu$, $\keta$ and $\kp$, and highlight this process for the Re470Pm2 and Re1700Pm2 simulations, but ultimately perform the numerical convergence study on all of our simulations. We estimate how the measured scales depend upon resolution by running all of our simulations at $\Nres = 18, 36, 72, 144$ and $288$, and some of them also at $\Nres = 576$ (indicated by $\dag$ in column (1) of \autoref{tab:simulations_setup}).

\begin{figure}
    \centering
    \includegraphics[width=\linewidth]{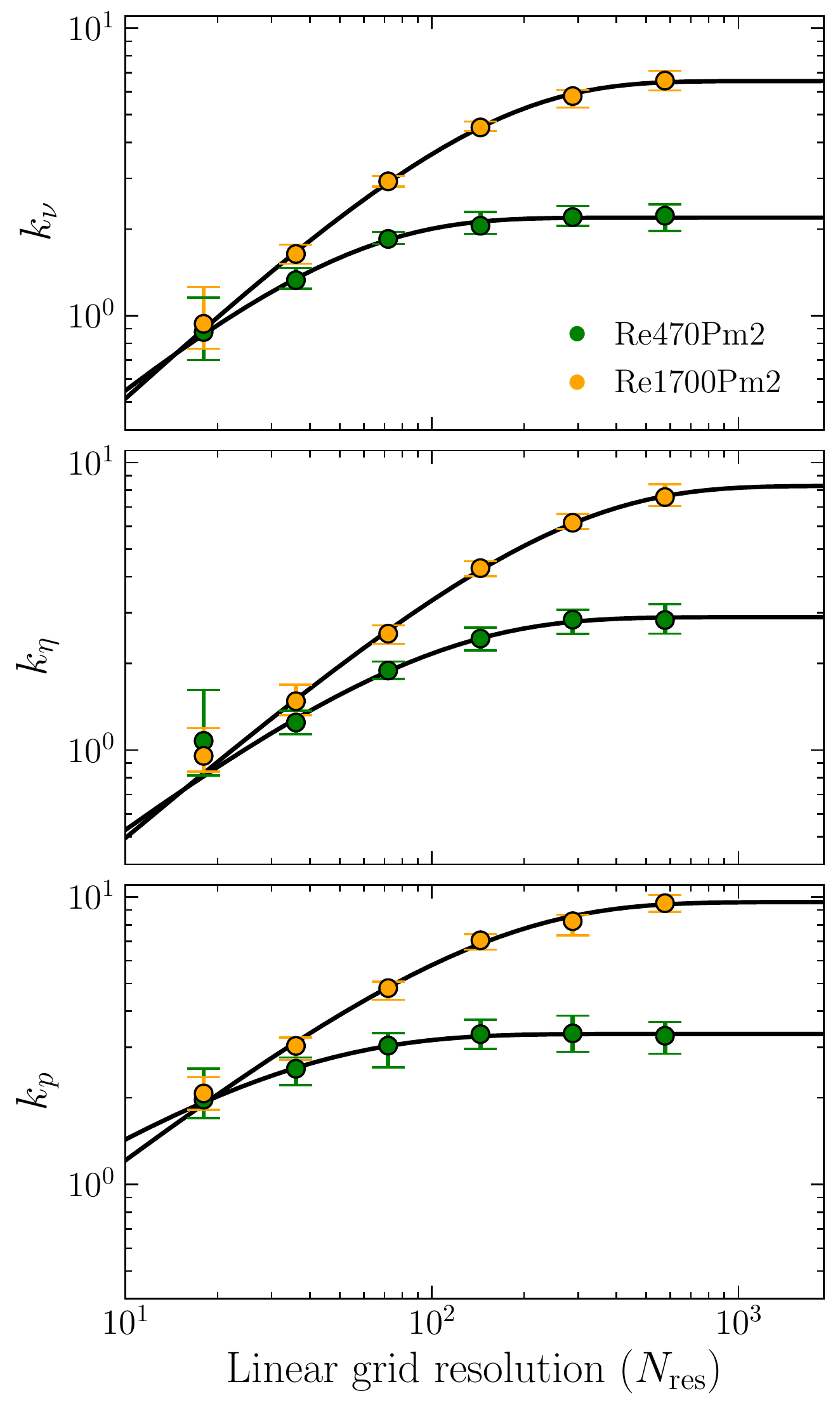}
    \caption{Measured dissipation wavenumbers, $\knu$ (top panel) and $\keta$ (middle panel), and the peak magnetic field scale (bottom panel), $\kp$, from our Re470Pm2 and Re1700Pm2 simulations, plotted against the linear grid resolution of the simulation, $\Nres$. We overlay our convergence model (see \autoref{eqn:convergence}) fitted for each of the simulations.}
    \label{fig:resolution}
\end{figure}

In \autoref{fig:resolution}, we show the measured $\knu$, $\keta$, and $\kp$ scales for our Re470Pm2 and Re1700Pm2 simulations against $\Nres$ of the simulations (in the top, middle, and bottom panels, respectively). As $\Nres$ increases, we find that the scales we measure move to higher $k$-values. However, the scales start to converge at around $\Nres \approx 100$ for Re470Pm2, and $\approx 300$ for Re1700Pm2. We quantify the rate of convergence, and measure the converged wavenumbers for $\knu$, $\keta$, and $\kp$ by fitting
\begin{equation}
    k_\text{scale}(\Nres) = k_\text{scale}(\Nres\to\infty) \cbrac{ 1 - \exp\left[ -(\Nres/\Nc)^r \right] } , \label{eqn:convergence}
\end{equation}
to each of the scales, for each of our simulations, where \mbox{$k_\text{scale}(\Nres\to\infty)$} $\equiv k_\text{scale}$ is the converged wavenumber for the simulation, $\Nc$ is the characteristic $\Nres$ where $k_\text{scale}$ starts to converge, and $r$ is the convergence rate.

We perform the convergence study for all of our simulations, and report the fitted convergence parameters in \autoref{eqn:convergence} for $\knu$, $\keta$, and $\kp$ in columns (2), (3), and (4) in \autoref{tab:converge_params}, respectively. We also report the converged wavenumbers $\knu$, $\keta$, and $\kp$ for all of our simulations, and report them in columns (12), (13), and (14) in \autoref{tab:simulations_setup}, respectively. From hereon, for the sake of simplicity, we will refer to the converged scales as $\knu$, $\keta$ and $\kp$, because we perform further analysis only with the converged scales.

\subsection{Measured dissipation scales vs.~theory} \label{subsec:relation}

Here we compare the measured and converged dissipation wavenumbers, $\knu$ and $\keta$, in our simulations with those predicted from current theories, namely $\knutheory$ (given by \autoref{eqn:knutheory}) and $\ketatheory$ (given by \autoref{eqn:ketatheory}).

\begin{figure*}
    \centering
    \includegraphics[width=\linewidth]{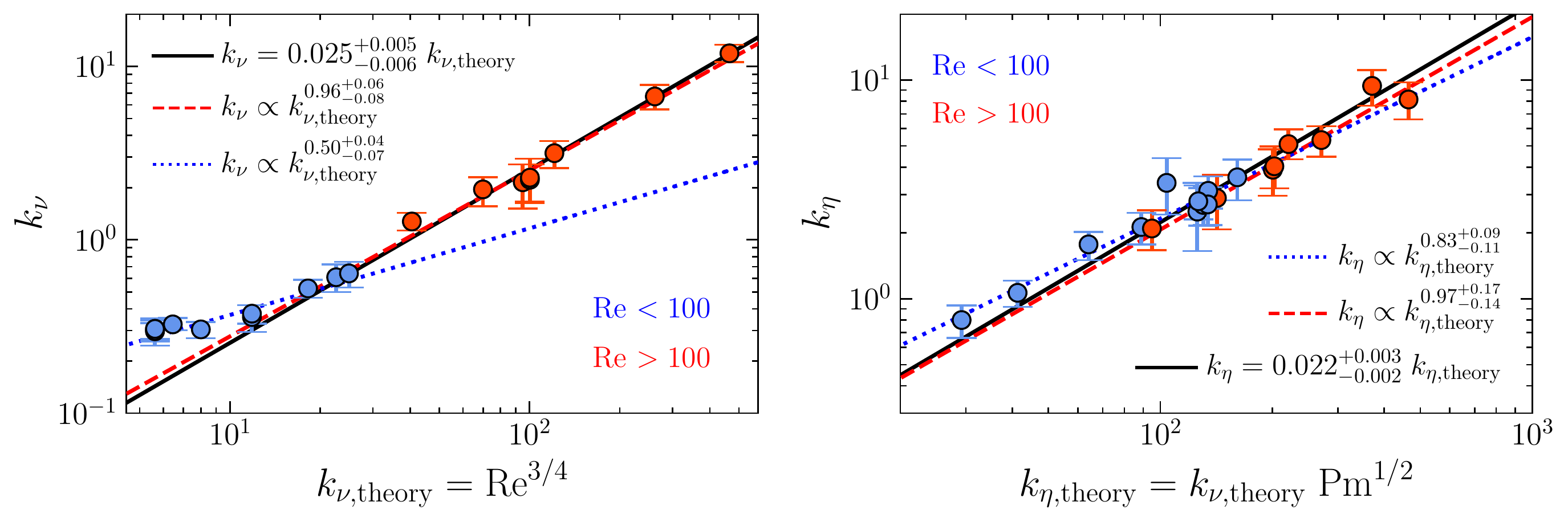}
    \caption{Measured dissipation wavenumbers, $\knu$ (left panel) and $\keta$ (right panel), compared with where theoretical relations predict the dissipation wavenumbers to be (see \autoref{eqn:knutheory} and \ref{eqn:ketatheory}). The measured and theoretical scales have a different relationship for simulations where $\Reyk < 100$ (blue) compared with $\Reyk > 100$ (red). We fit a power law to simulation points where $\Reyk < 100$ (blue dotted line), and also fit a power law model to simulations where $\Reyk > 100$ (red dashed line). We finally fit a linear model to the $\Reyk > 100$ points (black solid line).}
    \label{fig:scale_relation}
\end{figure*}

In \autoref{fig:scale_relation}, we show $\knu$ against $\knutheory$ (left panel), and $\keta$ against $\ketatheory$ (right panel). We separate points on the plot into two groups: (1) those scales measured from simulations where $\Reyk < 100$ (blue points), and (2) $\Reyk > 100$ (red points). The reason for this separation in $\Reyk$ will become clearer in the next section; for now, we will calculate statistics of the measured scales for $\Reyk < 100$ and $\Reyk > 100$, separately.

For simulations where $\Reyk < 100$, we measure that $\knu$ scales with $\knutheory$ as a power law with exponent $0.51_{-0.01}^{+0.04}$, and $\keta$ scales with $\ketatheory$ as a power law with exponent $0.83_{-0.09}^{+0.12}$. Conversely, for simulations where $\Reyk > 100$, we find a linear relationship (within the $1\sigma$ uncertainty) between the theoretical and measured dissipation wavenumbers for both $\knu$ and $\keta$. Thus, we conclude that the basic dependencies of $\knu$ on $\Reyk^{3/4}$, and $\keta$ on $\knutheory\, \Pranm^{1/2}$, follow the theoretical relations, but only if $\Reyk>100$. However, even if $\Reyk>100$, we find a significant shift between the measured and theoretical dissipation scales (quantified by a constant of proportionality). Fitting a linear model to the $\Reyk > 100$ points, we measure that the constant of proportionality between the measured and theoretical scales for $\knu$ and $\keta$ is $0.025^{+0.005}_{-0.006}$ and $0.021^{+0.002}_{-0.002}$, respectively. In summary, for $\Reyk > 100$, we find
\begin{equation}
    \knu = \left(0.025^{+0.005}_{-0.006}\right)\, k_\turb\, \Reyk^{3/4}, \label{eqn:knu_result}
\end{equation}
and
\begin{equation}
    \keta
    = \rbrac{0.88^{+0.21}_{-0.23}}\knu\, \Pranm^{1/2} . \label{eqn:keta_result}
\end{equation}
The utility of \autoref{eqn:knu_result} and \ref{eqn:keta_result} is that from $\Reyk$ and $\Reym$, they provide the exact viscous and resistive dissipation scales, $\knu$ and $\keta$, for subsonic MHD turbulence with $\Pranm \geq 1$.

\subsection{Dependence of the peak magnetic field scale on the turbulent and magnetic dissipation scales} \label{subsec:dependence}

\begin{figure*}
    \centering
    \includegraphics[width=\linewidth]{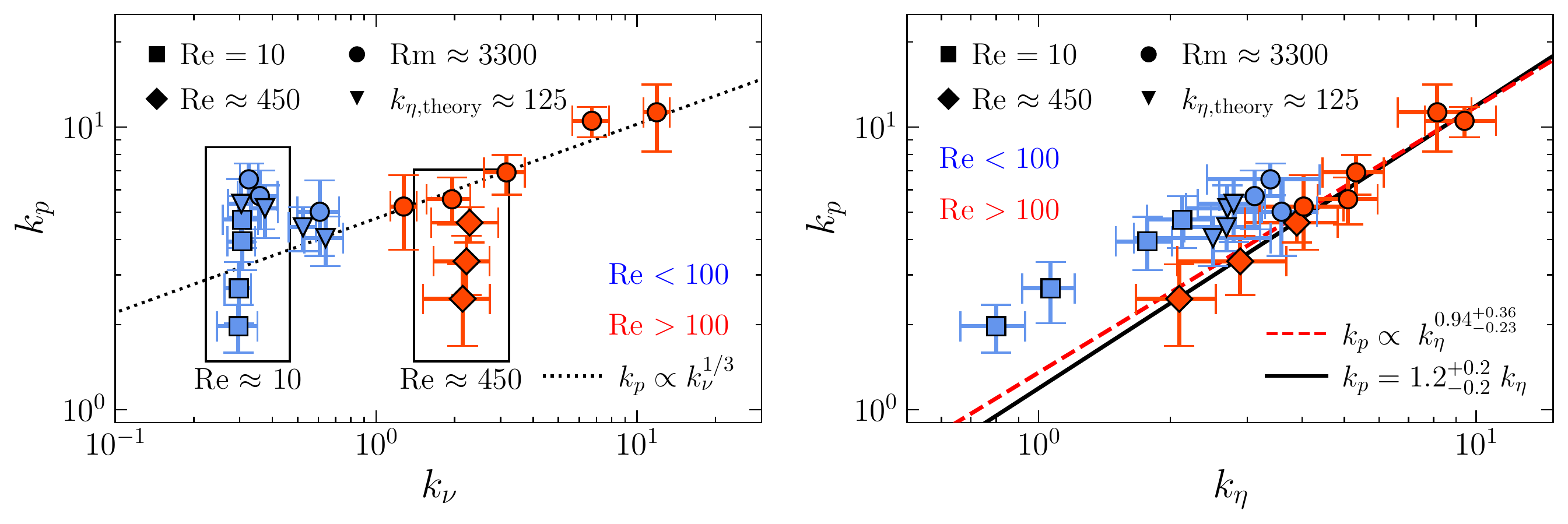}
    \caption{The measured peak magnetic field scale, $\kp$, plotted against the measured dissipation wavenumbers, $\knu$ (left panel) and $\keta$ (right panel). We plot scales measured in simulations where $\Reyk < 100$ in blue, and $\Reyk > 100$ in red. In the left panel we plot a power law, $\kp \propto \knu^{1/3}$ (dotted line), and put a box around scales where $\Reyk \approx 10$ and $450$ to highlight simulations where $\kp$ changed, but $\knu$ remained constant. In the right panel we fit both a power law (red dashed line) and linear model (black solid line) to simulations where $\Reyk > 100$. We report the measured power-law exponent, and the constant of proportionality for the power law and linear fits at the bottom right corner of the figure, respectively.}
    \label{fig:scale_dependance}
\end{figure*}

To determine the dependence of $\kp$ on $\knu$ and $\keta$, we show $\kp$ as a function of $\knu$ (left panel) and $\keta$ (right panel) in \autoref{fig:scale_dependance}. We first consider the relationship between $\kp$ and $\knu$ for simulations where $\Reyk$ has been fixed (see the boxes in the left-hand panel of \autoref{fig:scale_dependance}). Namely, there are two sets of simulations, $\Reyk \approx 10$ with $\Pranm = 27$--$250$, and $\Reyk \approx 450$ with $\Pranm = 1$--$4$. For these simulations we find that although $\knu$ is fixed in each set, there is an increase in $\kp$ with $\Pranm$. This suggests that $\knu$ cannot be the principle quantity that controls $\kp$. However, we also see that there are a few data points that seem to fall along a $\kp \propto \knu^{1/3}$ line. While this might suggest there could be a dependence of $\kp$ on $\knu$, we will see that this is in fact not a principle dependence.

In the right-hand panel of \autoref{fig:scale_dependance}, we see that an increase in $\keta$ always results in an increase in $\kp$. We find that there is a dichotomy in the relationship between $\kp$ and $\keta$, with $\kp$ scaling differently with $\keta$ for the $\Reyk < 100$, compared with the $\Reyk > 100$. Fitting a power law (red dashed line) to the $\Reyk > 100$ data points, we measure an exponent for the power law that captures unity within the $1\sigma$ uncertainty. We measure the linear relationship between $\kp$ and $\keta$ for the $\Reyk > 100$, shown by the solid black line.

We can now also understand that some of the simulations in the left-hand panel of \autoref{fig:scale_dependance} show a correlation between $\kp$ and $\knu$. Specifically, if $\kp \propto \keta$ is the fundamental relation, then it follows that $\kp \propto \knu\, \Pranm^{1/2} \propto \knu\, \Reym^{1/2}\Reyk^{-1/2} \propto \knu\, \Reym^{1/2}\knu^{-2/3} \propto \knu^{1/3}\, \Reym^{1/2}$. Thus, if $\Reym$ is fixed, then $\kp \propto \knu^{1/3}$, which is exactly what we observe in the left-hand panel of \autoref{fig:scale_dependance} for the $\Reym \approx \text{constant}$ subclass of models (see all rows in simulation suite $\Reym \approx 3300$ and the last two rows in $\ketatheory \approx 125$ in \autoref{tab:simulations_setup}). However, the scaling of $\kp$ on $\knu$ is simply a consequence of the fundamental underlying relation $\kp\propto\keta$.

This basic result agrees with current theories that predict ultimately the scale dependence of $\kp$ in the kinematic phase of the dynamo is on $\keta$ \citep{kazantsev1968enhancement, kulsrud1992spectrum, Vainshtein1992, schekochihin2002spectra, schekochihin2004simulations, brandenburg2005astrophysical, schober2015saturation, xu2016turbulent, mckee2020magnetic}. While the relation $\kp \propto \keta$ was anticipated in those theories, the constant of proportionality was less clear. Using our simulation suite we measure this constant of proportionality, by fitting a linear model (black solid line in the right-hand panel in \autoref{fig:scale_dependance}), and find
\begin{equation}
    \kp = \rbrac{1.2_{-0.2}^{+0.2}}\, \keta . \label{eqn:kp_result}
\end{equation}
Thus, we find that there is very little scale separation between $\kp$ and $\keta$, with the peak scale located close to the resistive scale. In \S\ref{subsec:kaz_exp} we discuss the origin of the proportionality constant, $\approx1.2$, and its relation to the properties of the magnetic energy spectrum.

\subsection{The Kazantsev exponent} \label{subsec:kaz_exp}

\begin{figure}
    \centering \includegraphics[width=\linewidth]{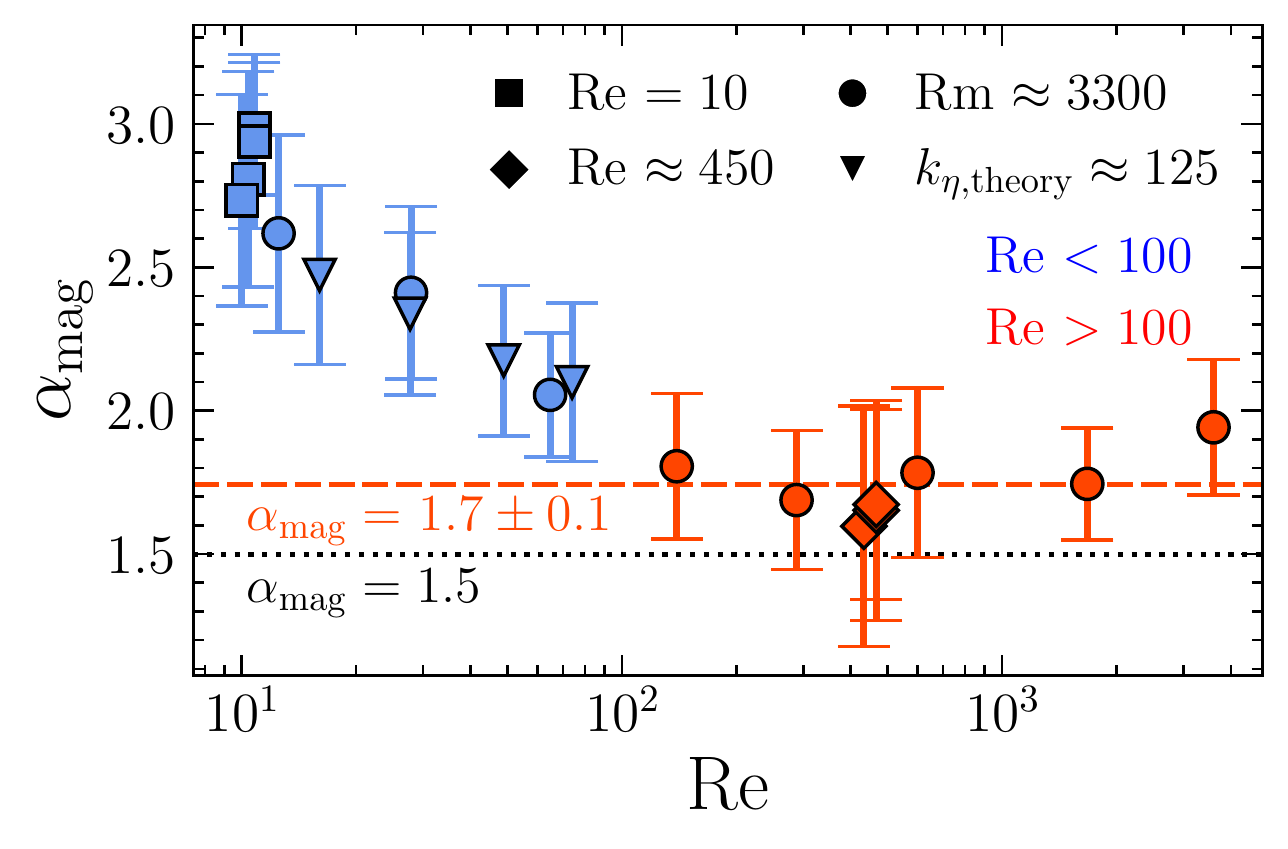}
    \caption{The magnetic power-law exponent, $\alpha_\tmag$ in \autoref{eqn:mag_spectra}, measured for each of our simulations (see \autoref{tab:simulations_setup}). We distinguish between simulations where $\Reyk < 100$ (blue) and $\Reyk > 100$ (red). We show the average $\alpha_\tmag$ for the $\Reyk > 100$ simulations as a red dashed line. We also plot a reference line indicating the Kazantsev exponent, $\alpha_\tmag = 3/2$ (black dotted line).}
    \label{fig:exponent}
\end{figure}

In \autoref{fig:exponent} we show the measured power-law exponent $\alpha_\tmag$ in the magnetic spectra of \autoref{eqn:mag_spectra}, from our fits in \S\ref{subsec:spectra_fits}, against $\Reyk$ for all our simulations. We find that for simulations where $\Reyk < 100$, $\alpha_\tmag$ increases with decreasing $\Reyk$. Conversely, for simulations where $\Reyk > 100$ we find that $\alpha_\tmag$ has reached a statistically steady value of $\alpha_\tmag = 1.7 \pm 0.1$. Recall that $\alpha_\tmag$, together with the ratio of the second-order Bessel functions in \autoref{eqn:kp_implicit}, sets the proportionality constant we measured in \autoref{eqn:kp_result}. Rearranging \autoref{eqn:kp_implicit} gives an implicit equation for the proportionality constant, $x = \alpha_\tmag\, K_0(x)/K_1(x)$, where $x \equiv \kp/\keta$. Solving this equation with $\alpha_\tmag = 1.7 \pm 0.1$ gives a proportionality constant of $1.3 \pm 0.1$, which agrees with our previous measurement in \autoref{eqn:kp_result} (see the right-hand panel of \autoref{fig:scale_dependance}). If $\alpha_\tmag = 3/2$, as suggested by Kazantsev's theory \citep{kazantsev1968enhancement, kulsrud1992spectrum}, then the proportionality constant would be $1.07$.

Throughout this study, we have found a dichotomy between $\Reyk < 100$ and $\Reyk > 100$. In summary, we found that the measured dissipation wavenumbers follow a scaling consistent with theoretical predictions, and that $\kp \propto \keta$. Here, we also find that $\alpha_\tmag \approx \text{constant}$. However, all of these properties are only seen if $\Reyk > 100$. By contrast, the scaling relations break down for $\Reyk < 100$ and $\alpha_\tmag$ starts to depend on $\Reyk$.

%% DISCUSSION
\section{Discussion} \label{sec:discussion}

\subsection{Limitations} \label{subsec:limitations}

In this paper we perform a systematic study wherein we measure $\knu$, $\keta$, and $\kp$ from MHD simulations (as described in \S\ref{subsec:spectra_fits}) in the kinematic phase of the turbulent dynamo, and determine that the principle dependence of $\kp$ in the subsonic, \mbox{$\Pranm \geq 1$} regime is on $\keta$. To isolate the dependence of $\kp$ on $\keta$, we vary the resistive scale in our simulations by changing $\eta$ in \autoref{MHD:induction}. While details of the magnetic dissipation processes can vary in nature, for example with ambipolar diffusion or Hall diffusion, where $\eta$ can be a function of the magnetic field, etc., here we only considered Ohmic dissipation with constant $\eta$ (i.e., we vary $\eta$ between different simulations, but $\eta$ is constant in space and time for a given simulation). While magnetic field dissipation may be more complex in nature, with $\eta$ depending on various processes, fixing $\eta$ allows us to set the magnetic Reynolds and Prandtl number, and therefore define a controlled value of $\keta$, for which we can measure the dependence on $\Reym$ and $\Pranm$.

% We overcome the numerical limitations of measuring the fundamental scales $\knu$, $\keta$, and $\kp$ from uniformly discretised simulations by performing a resolution study on each of these scales, for each of our simulations, which we discuss in detail in \S\ref{subsec:resolution}. Moreover, as discussed in \S\ref{sec:theory}, in the kinematic phase of the dynamo these scales are expected to be hierarchical, with $\ell_\turb > \ell_\nu > \ell_\eta$ (recall that $\ell_\text{scale} = 2\pi k_\text{scale}$). However, by simulating the turbulent dynamo we introduce an additional scale, namely the system-size scale $L$ (i.e., the box-length scale). Simultaneously resolving the wide range of scales present in our simulations, i.e., $L > \ell_\turb > \ell_\nu > \ell_\eta$, is not yet computationally feasible, especially when the scale separation between $\ell_\turb$ and $\ell_\nu$ (i.e., the inertial range), and/or $\ell_\nu$ and $\ell_\eta$ (i.e., the sub-viscous range) is large. Therefore, to access and explore a wide range of $\Pranm$ -- where in \S\ref{subsec:relation} we showed that $\Pranm^{1/2}$ defines the width of the sub-viscous range (as expected from \citet{schekochihin2002spectra}) -- we choose to drive the turbulent acceleration field with a profile such that the effective driving scale $\ell_\turb$ is equal to the box length, i.e., $\ell_\turb = L$ (see \S\ref{subsec:turb_driving} for details on the turbulent driving profile we use for all of our simulations).

Since it is not possible to do a large parameter study, where both the inertial and sub-viscous ranges are captured in each simulation, we choose to primarily focus on the sub-viscous range. To do this, for some of our simulations, we reduce the range of scales that the inertial range occupies by choosing \mbox{$\Reyk < 100$}. While the turbulent field produced by this driving is still random, due to the Ornstein-Uhlenbeck process, it is not entirely clear what effect the choice of the turbulent driving scale, namely \mbox{$\ell_\turb = L$}, has on the statistics of the turbulence. But if the correlation scale of the turbulence is approximately the driving scale, then setting the driving scale to be the entire box will reduce the independent spatial samples of the turbulence, thus making the statistics more sensitive to spatially intermittent events. \citet{schumacher2014small} showed that the velocity gradients (responsible for dissipative events in the velocity fields) in hydrodynamical turbulent simulations transition from Gaussian to intermittent at \mbox{$\Reyk \approx 100$}, where intermittent fluctuations of velocity gradients are characteristic of fully-developed turbulence. It is not clear what properties differentiate our \mbox{$\Reyk < 100$} and \mbox{$\Reyk > 100$} simulations, but we hypothesise that it could be related to the intermittency of the velocity field gradients in the turbulence, as suggested by \citet{schumacher2014small} (see \aref{app:vel_gradients} for more details on the measured higher-order moments of the velocity gradients from our simulations).

In this study, we consider plasmas where \mbox{$\Pranm \geq 1$}, and while we solve the compressible MHD equations, we fix \mbox{$\Mach \approx 0.3$}. The astrophysical relevance of this regime is discussed in detail in the next subsection, but in many astrophysical systems, velocity fluctuations can become highly-compressible and supersonic. For example in regions of the cold interstellar medium, turbulence is compressible with \mbox{$\Mach \sim 10$} \citep{elmegreen2004interstellar, federrath2016link, Beattie2019_mach_number_fractal}. It has been shown that compressibility affects the dynamo efficiency \citep[see for example][]{haugen2004mach, federrath2011mach, federrath2014turbulent, seta2021saturation}, thus further numerical experiments are necessary to understand the dependence of $\kp$ on $\knu$ and $\keta$ when the turbulence is supersonic. Moreover, it is also important to study how the dependence of $\kp$ on $\keta$ changes when \mbox{$\Pranm < 1$}, such as in the Sun's convective zone (also in planets, and in protostellar discs, where \mbox{$\Pranm \ll 1$}), because if \mbox{$\Pranm < 1$}, then \mbox{$\keta < \knu$}, and it is unlikely that the scaling relations established here continue to hold.

Finally, we only study the kinematic growth phase of the dynamo. While the kinematic phase of the turbulent dynamo is responsible for amplifying weak magnetic seed fields by many orders of magnitude, inevitably, the magnetic field becomes strong enough to resist further amplification and exerts a back-reaction (via the Lorentz force) on the turbulent velocity field.  This marks the transition from the kinematic phase of the dynamo to the nonlinear phase. In future works, we wish to determine whether an evolution of $\kp$ is present in the early kinematic phase, as suggested by \citet{schekochihin2002spectra, xu2016turbulent, mckee2020magnetic}, and whether $\kp$ shifts from $\sim \keta$ to larger scales (i.e., lower $k$) as the dynamo transitions through the nonlinear phase to the saturated phase \citep{xu2016turbulent, mckee2020magnetic, galishnikova2022tearing}. This will require very high $\Pranm$ in order to maximise scale separation between $\knu$ and $\keta$, which is challenging, but may be possible with future very-high-resolution simulations. Regardless of the dynamics that take place in the nonlinear and saturated phases of the dynamo, it is clear that the kinematic phase sets the initial conditions conditions for the strength and structure of the magnetic field, where we have showed that \mbox{$\kp = 1.2_{-0.2}^{+0.2} \keta$} in the kinematic phase.

\subsection{Implications} \label{subsec:implications}

Primordial magnetic fields must have been amplified by turbulent dynamos to the dynamically significant field strengths that we observe today (see \S\ref{sec:theory} and references therein). We have explored turbulent dynamos in the \mbox{$\Pranm \geq 1$} and incompressible regime, which holds application for magnetic fields in the early Universe, and more broadly for hot, low-density astrophysical plasmas, such as in the warm interstellar medium, accretion discs, protogalaxies, and the intracluster gas in galaxy clusters \citep[see for example][]{kulsrud1992spectrum, kulsrud1999critical, schekochihin2002model, schekochihin2004simulations, shukurov2004introduction, vazza2018resolved, gent2021small}.

Using numerical simulations, we have quantified the distribution of magnetic energy as a function of scale \citep[modelled by \autoref{eqn:mag_spectra}; see][]{kulsrud1992spectrum}, which tells us where the magnetic energy is the strongest (i.e., the peak scale $\kp$) and where it dissipates ($\keta$). We have determined how exactly $\kp$ and $\keta$ depend on the hydrodynamic and magnetic Reynolds numbers ($\Reyk$ and $\Reym$) of any turbulent, magnetised system (\autoref{eqn:knu_result} and \ref{eqn:keta_result}) during the phase of the dynamo where magnetic field energy amplifies most significantly (kinematic phase of the turbulent dynamo). The implications this holds is that for a given $\Reyk$ and $\Reym$, one can directly calculate $\knu$ (viscous wavenumber) and $\keta$, and determine the entire spectrum of magnetic energy via \autoref{eqn:mag_spectra}.

In the astrophysical environments mentioned above, $\Reyk$ and $\Reym$ vary over many orders of magnitude \citep{schekochihin2007fluctuation}, where our results allow us to derive the distribution of magnetic energy in the kinematic phase of the dynamo taking place in these systems. For example, considering star formation in primordial halos, \citet{schober2012small} and \citet{nakauchi2021ionization} calculate $\Reyk$ and $\Reym$, and using these numbers, our results provide the scale-dependent magnetic energy during the kinematic phase of a turbulent dynamo. Doing this, our results imply that the magnetic energy is concentrated on scales much smaller than the size of primordial mini-halos, i.e., the field is strongest in the dense regions where accretion discs and stars form. At later stages of the dynamo (nonlinear phase of the dynamo; see the discussion at the end of \S\ref{subsec:limitations}), these fields can provide support against collapse and suppress fragmentation of the first-star discs, thereby reducing the number of low-mass stars that formed \citep{sharda2020importance, sharda2021magnetic, stacy2022magnetic}. These small-scale magnetic fields may also give rise to protostellar outflows and jets \citep{machida2006first, machida2019first}.

% Although turbulent dynamo amplified magnetic fields become concentrated at smaller scales compared with the turbulent velocity fields that drive the dynamo process, they helped shape the Universe we live in today, and could help explain structure formation in the early Universe \citep{wagstaff2014magnetic}.

% Here we have also tested and confirmed that the theoretical scaling relations, i.e., $\knutheory$ and $\ketatheory$ (given by \autoref{eqn:knutheory} and \ref{eqn:ketatheory}, respectively), that are frequently used in the literature \citep[for example in][]{schekochihin2002spectra, schekochihin2004critical, schekochihin2004simulations, schekochihin2007fluctuation, rincon2016turbulent, xu2016turbulent, mckee2020magnetic}, are functionally correct. Further, we have provided quantitative modifications to these relations, and determined the complete form of these relations (see \autoref{eqn:knu_result} and \ref{eqn:keta_result}, respectively).

Another application of our main results (\autoref{eqn:knu_result} and \ref{eqn:keta_result}) would be to determine the effective kinematic and magnetic Reynolds numbers in ideal, incompressible MHD simulations. Since in ideal MHD simulations $\Reyk$ and $\Reym$ are not controlled by physical dissipation, but rather set by numerical viscosity and resistivity, one does not know the exact values of $\Reyk$ and $\Reym$. Fitting \autoref{eqn:kin_spectra} and \ref{eqn:mag_spectra} to the kinetic and magnetic power spectra obtained in ideal MHD simulations, one is able to extract $\knu$ and $\keta$, and by inverting relations, \autoref{eqn:knu_result} and \ref{eqn:keta_result}, one can directly calculate the effective $\Reyk$ and $\Reym$ for the simulations.

%Finally, while simulations currently only have access to the low-$\Reyk$ and low-$\Reym$ regime, we have established \autoref{eqn:knu_result} and \ref{eqn:keta_result} over a large range of $\Reyk$ and $\Reym$, thus, it should be possible to extrapolate our results in this study into the high-$\Reyk$ and $\Pranm$ regime with high confidence.

%% CONCLUSION
\section{Summary and conclusions} \label{sec:summary}

We have used direct numerical simulations of MHD turbulent dynamo action to measure the viscous scale ($\knu$), the resistive scale ($\keta$), and the peak magnetic field scale ($\kp$), in $104$~simulations with hydrodynamic Reynolds numbers $10 \leq \Reyk \leq 3600$, and magnetic Prandtl numbers $1\leq \Pranm\leq 260$ (see \autoref{tab:simulations_setup} and \S\ref{subsec:initalise} for details of the simulations). There has been some disagreement in the literature about whether $\kp$ should be concentrated at $\knu$ or $\keta$ (see \S\ref{sec:theory} and references therein). Here we determine the fundamental dependence of $\kp$ for $\Pranm \geq 1$, which we find is on $\keta$, and not on $\knu$. However, we also demonstrate that $\keta \propto \knu\, \Pranm^{1/2} \propto \knu^{1/3}\, \Reym^{1/2}$, following theoretical predictions, and thus, in a limited parameter set where $\Pranm$ or $\Reym$ had been fixed, the principle dependence of $\kp$ on $\keta$ could have been mistaken for a principle dependence on $\knu$.

In the following, we summarise our study in item format:
\begin{itemize}
    \item We first confirm the exponential amplification of the magnetic field during the kinematic phase of the dynamo (see \autoref{fig:energy_evol}), with growth rates, $\grate$ and magnetic-to-kinetic energy saturation levels $\satlevel$, that depend upon $\Reyk$. We find general agreement between our measurements of $\grate$ and $\satlevel$, for our simulations, compared with measurements in previous analytic and numerical works. \\
    
    \item In \autoref{fig:energy_slice} we show two-dimensional slices of kinetic and magnetic energy, and observe smaller-scale structures for large $\Reyk$ compared to small $\Reyk$. \\
    
    \item We quantify the size of these field structures by studying the (time-averaged) kinetic and magnetic power spectra, $\Pvel$ and $\Pmag$, respectively, shown in \autoref{fig:spectra}. We fit \autoref{eqn:kin_spectra} and \autoref{eqn:mag_spectra} to $\Pvel$ and $\Pmag$, respectively, allowing us to measure $\knu$, $\keta$, and $\kp$. We make sure that these scale measurements are numerically converged (see \autoref{fig:resolution}). \\
    
    \item With robust measurements of $\knu$, $\keta$, and $\kp$ for all of our simulations, we find that for $\Reyk > 100$, there is excellent agreement in the scaling of the dissipation wavenumbers we measure from our simulations, $\knu$ and $\keta$, and the theoretical relations in the literature, $\knutheory = k_\turb \Reyk^{3/4}$ and $\ketatheory = \knutheory \Pranm^{1/2}$, where $k_\turb$ is the turbulence driving scale (see \autoref{fig:scale_relation}). However, we find a significant offset by a constant factor between $\knu$ and $\knutheory$, and between $\keta$ and $\ketatheory$, respectively. We measure these two constants of proportionality, and determine the overall dissipation scale relations, $\knu=(0.025^{+0.005}_{-0.006})\, k_\turb\, \Reyk^{3/4}$ and $\keta=(0.022_{-0.002}^{+0.003})\, k_\turb\, \Reyk^{3/4}\, \Pranm^{1/2}=(0.88^{+0.21}_{-0.23})\, \knu\, \Pranm^{1/2}$. \\
    
    \item We measure that $\kp$ scales linearly with $\keta$ (see the right-hand panel in \autoref{fig:scale_dependance}). For simulations with $\Reyk > 100$, the relationship is $\kp = (1.2_{-0.2}^{+0.2}) \, \keta$. We find that the $1.2$ constant of proportionality in this relationship is related to the power-law exponent $\alpha_\tmag$ of the magnetic power spectrum. For $\Reyk > 100$, we find $\alpha_\tmag = 1.7 \pm 0.1$ (see \autoref{fig:exponent}), slightly larger, but close to the theoretical Kazantsev exponent of $3/2$. \\
    
    \item Throughout this study, we find that the fundamental properties of turbulent dynamo amplification break down for $\Reyk < 100$. Conversely, we see good agreement between our simulations and predictions of turbulent dynamo theory for $\Reyk > 100$. In this regime, our simulations have allowed us to determine the proportionality constants in theoretical relations of the turbulent dynamo, which so far remained largely unconstrained. We conclude that $\Reyk > 100$ is required for bonafide turbulent dynamo amplification, which is most likely a consequence of $\Reyk \gtrsim 100$ being the minimum requirement for fully-developed turbulent flow \citep[see also work by][]{frisch1995turbulence, schumacher2014small}. We show in \aref{app:vel_gradients} that the universal small-scale velocity gradient statistics of turbulence changes around $\Reyk \approx 100$, which is in good agreement with results from \citet{schumacher2014small}.
\end{itemize}

%% ACKNOWLEDGEMENTS
\section*{Acknowledgements}
We thank the anonymous referee for their useful comments, which helped to improve this work. N.~K. acknowledges funding from the Research School of Astronomy and Astrophysics, ANU, through the Bok Honours scholarship. J.~R.~B. acknowledges funding from the ANU, specifically the Deakin PhD and Dean's Higher Degree Research (theoretical physics) Scholarships and the Australian Government via the Australian Government Research Training Program Fee-Offset Scholarship. C.~F.~acknowledges funding provided by the Australian Research Council (Future Fellowship FT180100495), and the Australia-Germany Joint Research Cooperation Scheme (UA-DAAD). We further acknowledge high-performance computing resources provided by the Australian National Computational Infrastructure (grant~ek9) in the framework of the National Computational Merit Allocation Scheme and the ANU Merit Allocation Scheme, and by the Leibniz Rechenzentrum and the Gauss Centre for Supercomputing (grants~pr32lo and pn73fi and GCS Large-scale projects~10391 and 22542). The simulation software \textsc{flash} was in part developed by the DOE-supported Flash Center for Computational Science at the University of Chicago.

%% DATA AVAILABILITY
\section*{Data availability}
The simulation data underlying this paper will be shared on reasonable request to the corresponding author.

%% REFERENCES
\bibliographystyle{mnras}
\bibliography{refs}

%% APPENDICES
\appendix
\section{The Kolmogorov exponent} \label{app:kolmog_exp}

The limited scaling ranges in the turbulent kinetic energy spectra of our simulations (see the top panel of \autoref{fig:spectra}) do not allow us to constrain $\alpha_\tkin$ well. The main purpose of the present simulations, however, are not to measure the power-law scaling exponents of the turbulence \citep[which requires much higher resolution; see e.g., ][]{federrath2013universality, federrath2021sonic}; instead, we only want our simulations to capture the dissipation scales and sub-viscous range well. To confirm that the exact value of $\alpha_\tkin$ does not influence our main results, here, we explore the effects of fixing $\alpha_\tkin = -5/3$ in the \autoref{eqn:kin_spectra} model, which is the expected exponent for \citet{Kolmogorov1941} turbulence \citep[for simplicity, we ignore intermittency effects, which would introduce $\lesssim10\%$ corrections to the $-5/3$ scaling exponent; see][]{she1994universal, boldyrev2000geometric, schmidt2008scaling}, while still fitting for $A_\tkin$ and $\knu$. As in the main part of the study, we only fit to time realisations of the simulation where $10^{-6} \leq E_\tmag/E_\tkin \leq 10^{-2}$, and perform the scale convergence test discussed in \S\ref{subsec:resolution} on the time-averaged $\knu$ values for each of our simulations.

\begin{figure}
    \centering
    \includegraphics[width=\linewidth]{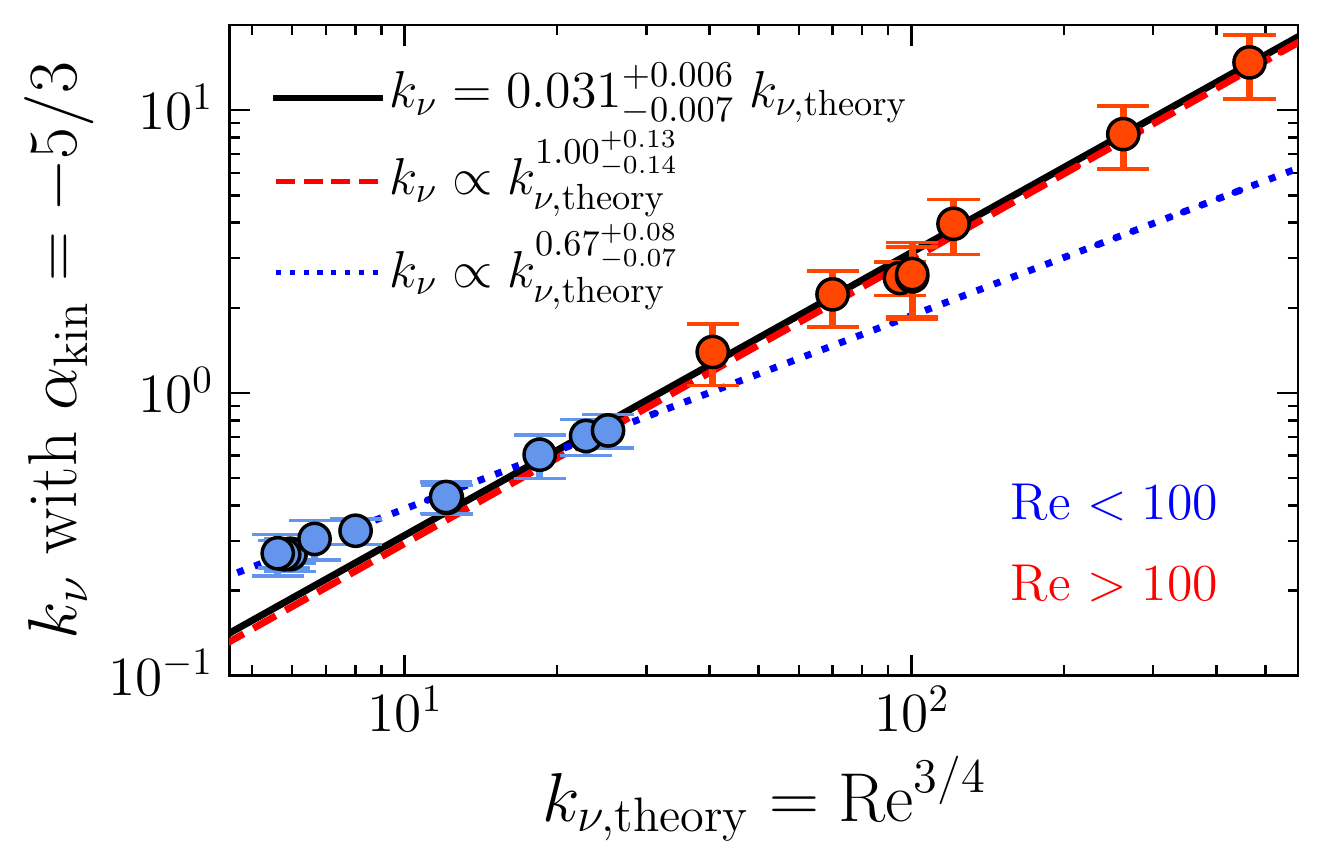}
    \caption{The same as in the left panel of \autoref{fig:scale_relation} but with the viscous dissipation wavenumber $\knu$ measured from the velocity spectra, \autoref{eqn:kin_spectra}, with a fixed $\alpha_\tkin = -5/3$.}
    \label{fig:k_nu_relation_fixed}
\end{figure}

In \autoref{fig:k_nu_relation_fixed}, we compare the measured and converged $\knu$ scales from \autoref{eqn:kin_spectra}, with $\alpha_\tkin = -5/3$, for all our simulations against the scales predicted by $\knutheory$ (given by \autoref{eqn:knutheory}). As in \autoref{fig:scale_relation}, we separate points into two groups: (1) those where the simulation $\Reyk < 100$, and (2) those where $\Reyk > 100$. For simulations where $\Reyk < 100$, we measure that $\knu$ scales with $\knutheory$ as a power law with exponent $0.67_{-0.14}^{+0.13}$ (see blue dotted line). For the $\Reyk > 100$ data points, we find that fitting a power law (red dashed line) gives a linear relationship (within the $1\sigma$ uncertainty) between $\knu$ and $\knutheory$. From this, we conclude that the basic dependence of $\knu$ on $\Reyk^{3/4}$ still holds true for the $\Reyk > 100$ data (as we had found in the left panel of \autoref{fig:scale_relation} in the main text). Moreover, by fitting a linear model (black line) to the $\Reyk > 100$ data, we measure a constant of proportionality of $0.031_{-0.007}^{+0.006}$. The constant of proportionality is slightly higher compared with the constant of proportionality in the main text,  $0.025_{-0.006}^{+0.005}$, where $\alpha_\tkin$ was a free parameter. However, both of the constants of proportionality overlap within their $1\sigma$ uncertainties. Thus, we conclude that fixing $\alpha_\tkin = -5/3$ does not significantly change the $\knu$ scales we measure from \autoref{eqn:kin_spectra}.

\begin{figure}
    \centering
    \includegraphics[width=\linewidth]{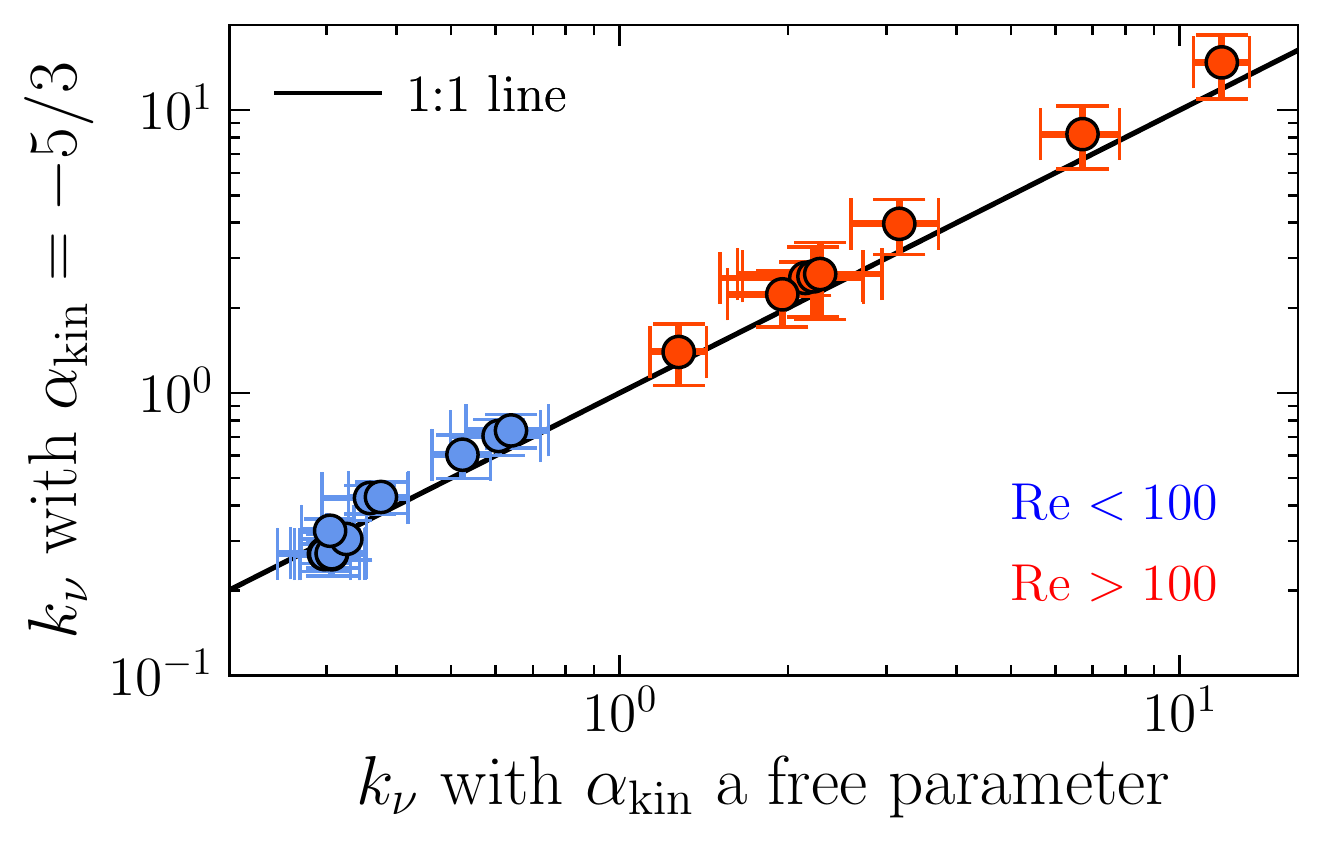}
    \caption{The viscous dissipation wavenumbers $\knu$ measured from fitting \autoref{eqn:kin_spectra} with $\alpha_\tkin = -5/3$ fixed (on the y-axis), compared to fitting \autoref{eqn:kin_spectra} with $\alpha_\tkin$ as a free parameter (on the x-axis), with a 1:1 line in solid black.}
    \label{fig:k_nu_comparison}
\end{figure}

The minor influence of $\alpha_\tkin$ on the $\knu$ scales that we measure is also reflected in \autoref{fig:k_nu_comparison}.  We plot $\knu$ measured from fitting \autoref{eqn:kin_spectra} with $\alpha_\tkin = -5/3$ fixed (on the y-axis), and compare these scales with $\knu$ measured by fitting \autoref{eqn:kin_spectra} with $\alpha_\tkin$ as a free parameter (on the x-axis), as in the main part of the study. We find that the measured $\knu$ from the two models follow a $1:1$ line, with $\knu$ slightly higher (by $\sim 20\%$) in the model where $\alpha_\tkin = -5/3$. However, both methods give $\knu$ that agrees well with one another within the $1\sigma$ uncertainty (see column~10 in \autoref{tab:simulations_setup}). Thus, the fact that the present simulations do not provide accurate measurements or strong constraints of $\alpha_\tkin$ does not have any significant influence on the main results of this study.

\section{Fit parameters of the numerical convergence study}
\autoref{tab:converge_params} lists the fit parameters of the numerical scale convergence study from \S\ref{subsec:resolution}.

\begin{table*}
    \centering
    \renewcommand{\arraystretch}{0.85}
    \setlength{\tabcolsep}{3.5pt}
    \caption{Derived scale convergence parameters.}
    \label{tab:converge_params}
    \begin{tabular}{l cc l cc l cc}
        \hline\hline
            Simulation
                & \multicolumn{2}{c}{$\knu$}
                && \multicolumn{2}{c}{$\keta$}
                && \multicolumn{2}{c}{$\kp$} \\
            \cline{2-3} \cline{5-6} \cline{8-9} \\[-1.5ex]
            ID
                &  $\Nc$ & $r$
                && $\Nc$ & $r$
                && $\Nc$ & $r$ \\
            (1)
                & \multicolumn{2}{c}{(2)}
                && \multicolumn{2}{c}{(3)}
                && \multicolumn{2}{c}{(4)} \\
        \hline\hline
        \multicolumn{9}{c}{$\Reyk = 10$} \\
        \hline
        Re10Pm27
            & $7 \pm 6$ & $0.5 \pm 0.4$
            && $20 \pm 10$ & $0.5 \pm 0.5$
            && $15 \pm 1$ & $0.56 \pm 0.02$ \\
        Re10Pm54
            & $9 \pm 6$ & $0.6 \pm 0.5$
            && $30 \pm 10$ & $0.5 \pm 0.3$
            && $27 \pm 2$ & $0.50 \pm 0.04$ \\
        Re10Pm130
            & $11 \pm 5$ & $0.5 \pm 0.4$
            && $80 \pm 70$ & $0.5 \pm 0.3$
            && $80 \pm 20$ & $0.50 \pm 0.07$ \\
        Re10Pm250
            & $11 \pm 6$ & $0.8 \pm 0.7$
            && $90 \pm 50$ & $0.7 \pm 0.2$
            && $83 \pm 4$ & $0.59 \pm 0.02$ \\
        \hline
        \multicolumn{9}{c}{$\Reyk \approx 450$} \\
        \hline
        Re430Pm1
            & $39 \pm 8$ & $1.0 \pm 0.4$
            && $40 \pm 40$ & $0.6 \pm 0.6$
            && $9 \pm 7$ & $0.9 \pm 0.7$ \\
        Re470Pm2
            & $38 \pm 6$ & $0.9 \pm 0.3$
            && $70 \pm 20$ & $0.8 \pm 0.2$
            && $22 \pm 1$ & $0.73 \pm 0.06$ \\
        Re470Pm4
            & $36 \pm 8$ & $0.9 \pm 0.3$
            && $(1.1 \pm 0.5)\times 10^2$ & $0.9 \pm 0.2$
            && $35 \pm 2$ & $0.84 \pm 0.06$ \\
        \hline
        \multicolumn{9}{c}{$\Reym \approx 3300$} \\
        \hline
        Re3600Pm1
            & $(2 \pm 1)\times 10^2 $ & $1.0 \pm 0.1$
            && $(2 \pm 1)\times 10^2 $ & $1.0 \pm 0.2$
            && $(2 \pm 1)\times 10^2 $ & $0.77 \pm 0.07$ \\
        Re1700Pm2
            & $(1.2 \pm 0.2)\times 10^2$ & $1.0 \pm 0.1$
            && $(2.1 \pm 0.6)\times 10^2$ & $0.9 \pm 0.1$
            && $(1.1 \pm 0.1)\times 10^2$ & $0.84 \pm 0.06$ \\
        Re600Pm5
            & $60 \pm 10$ & $0.9 \pm 0.2$
            && $(1.5 \pm 0.8)\times 10^2$ & $0.9 \pm 0.2$
            && $90 \pm 20$ & $0.75 \pm 0.06$ \\
        Re290Pm10
            & $36 \pm 7$ & $0.8 \pm 0.3$
            && $(1.4 \pm 0.4)\times 10^2$ & $0.9 \pm 0.1$
            && $63 \pm 9$ & $0.80 \pm 0.08$ \\
        Re140Pm25
            & $20 \pm 10$ & $0.5 \pm 0.5$
            && $(2 \pm 2)\times 10^2$ & $0.8 \pm 0.2$
            && $70 \pm 20$ & $0.7 \pm 0.1$ \\
        Re64Pm50
            & $5 \pm 30$ & $0.6 \pm 2.4$
            && $(1.2 \pm 0.9)\times 10^2$ & $0.8 \pm 0.2$
            && $50 \pm 9$ & $0.73 \pm 0.09$ \\
        Re27Pm128
            & $13 \pm 9$ & $0.8 \pm 0.8$
            && $(1.4 \pm 0.9)\times 10^2$ & $0.6 \pm 0.1$
            && $84 \pm 5$ & $0.62 \pm 0.02$ \\
        Re12Pm260
            & $13 \pm 4$ & $0.8 \pm 0.6$
            && $(6 \pm 3)\times 10^2$ & $0.5 \pm 0.2$
            && $(2.2 \pm 0.9)\times 10^2$ & $0.50 \pm 0.03$ \\
        \hline
        \multicolumn{9}{c}{$\ketatheory \approx 125$} \\
        \hline
        Re73Pm25
            & $8 \pm 10$ & $0.5 \pm 0.6$
            && $70 \pm 40$ & $0.8 \pm 0.2$
            && $36 \pm 2$ & $0.77 \pm 0.04$ \\
        Re48Pm51
            & $8 \pm 23$ & $0.7 \pm 2.1$
            && $90 \pm 70$ & $0.7 \pm 0.2$
            && $42 \pm 3$ & $0.71 \pm 0.03$ \\
        Re25Pm140
            & $13 \pm 9$ & $1 \pm 2$
            && $(1.1 \pm 0.6)\times 10^3$ & $0.5 \pm 0.2$
            && $69 \pm 2$ & $0.60 \pm 0.05$ \\
        Re16Pm250
            & $16 \pm 4$ & $1.1 \pm 0.8$
            && $(1.4 \pm 0.7)\times 10^2$ & $0.6 \pm 0.2$
            && $(1.2 \pm 0.9)\times 10^2$ & $0.6 \pm 0.1$ \\
        \hline\hline
    \end{tabular}
    \begin{tablenotes}
        \item \textit{Note}: All parameters are derived from fits of \autoref{eqn:convergence} to the time-averaged scales reported for each simulation ID \textbf{(1)} in \autoref{tab:simulations_setup}.
        Next we report the characteristic grid resolution, $\Nc$, and the convergence rate, $r$, measured for the viscous dissipation wavenumber, $\knu$ \textbf{(2)}, the resistive scale, $\keta$ \textbf{(3)}, and for the peak magnetic field scale, $\kp$ \textbf{(4)}, for each simulation.
    \end{tablenotes}
\end{table*}

\section{Non-Gaussian components of the velocity gradients} \label{app:vel_gradients}

\begin{figure}
    \centering
    \includegraphics[width=\linewidth]{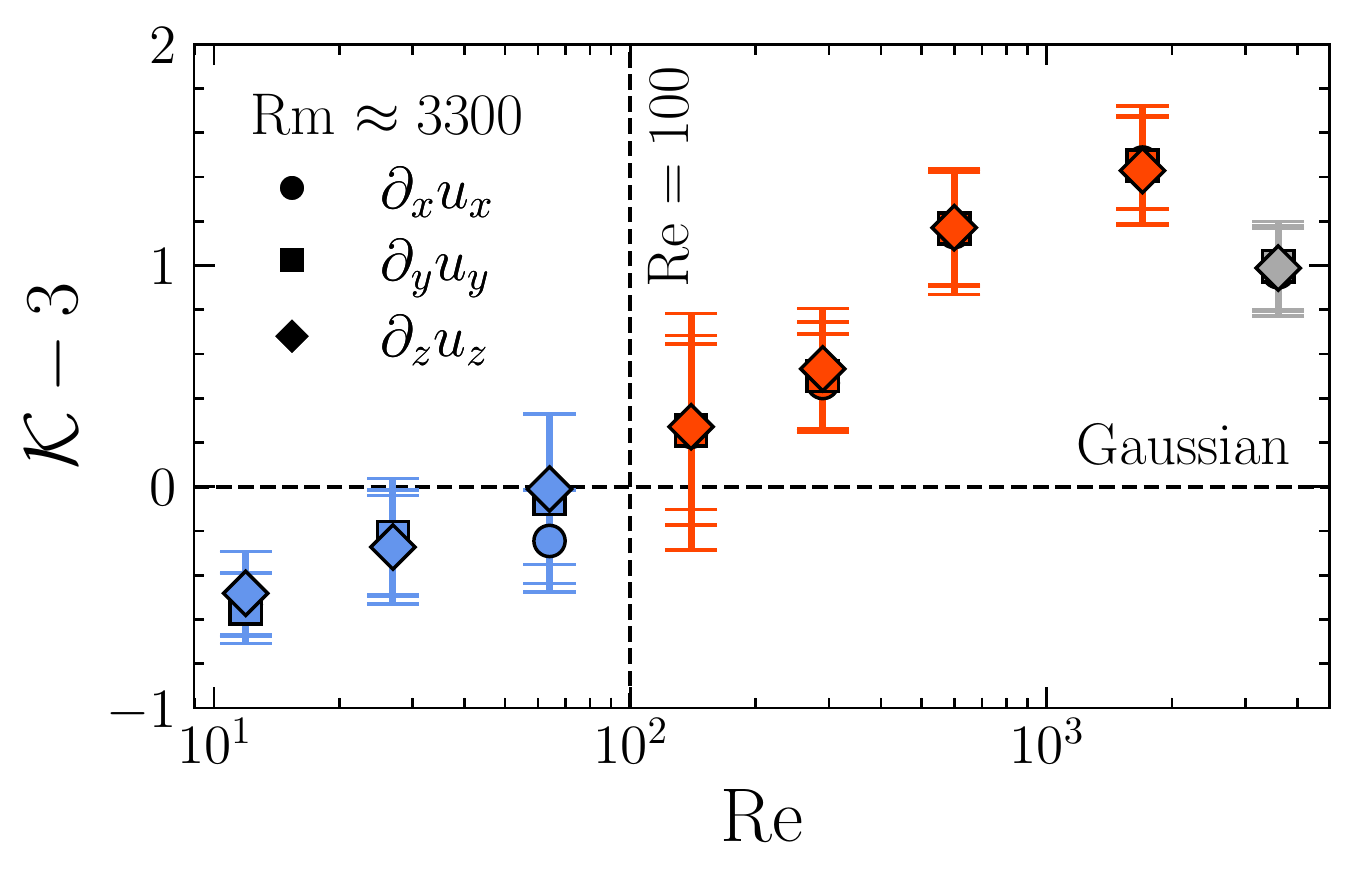}
    \caption{The excess kurtosis, $\mathcal{K} - 3$ (see \autoref{eqn:kurtosis} for the definition of $\mathcal{K}$), calculated for velocity gradients, $\partial_i u_i$, measured and time-averaged over $2 \teddy$ in the kinematic phase of the dynamo for each simulation in the $\Reym \approx 3300$ simulation suite with $12 \lesssim \Reyk \lesssim 3600$, at $\Nres = 288$. As expected, the kurtosis values for each simulation are similar in all three directions. Here, we distinguish between simulations where $\Reyk < 100$ (blue) and $\Reyk > 100$ (red), as in previous Figures, but plot Re3600Pm1 in grey to indicate that numerical viscosity will influence $\partial_i u_i$ in this simulation at $\Nres = 288$ (see \autoref{foot:high_Re_res} for details). We also indicate the transitions $\mathcal{K} - 3 = 0$, which corresponds to Gaussian velocity gradients (horizontal, black-dashed line), and $\Reyk = 100$, about which we measure that the velocity gradient statistics transition from Gaussian to super-Gaussian (vertical, black-dashed line).}
    \label{fig:kurtosis}
\end{figure}

Here we provide additional evidence for our hypothesis that the intermittency of the velocity gradient field (related to the dissipative structures in the velocity) can be responsible for the $\Reyk$ dichotomy (see Figures \ref{fig:scale_relation}, \ref{fig:scale_dependance}, and \ref{fig:exponent}). Following \citet{schumacher2014small}, we measure the kurtosis of the diagonal elements for the velocity gradient tensor, $\partial_i u_i$,
\begin{equation}
    \mathcal{K} = \frac{\langle(\partial_i u_i)^4\rangle_{\mathcal{V}}}{\langle(\partial_i u_i)^2\rangle_{\mathcal{V}}^2}
    \label{eqn:kurtosis}
\end{equation}
for each of the simulations in the $\Reym \approx 3300$ suite\footnote{
    \label{foot:high_Re_res}
    Note that, without the use of the numerical resolution correction method that we outline in \S\ref{subsec:resolution}, the Re3600Pm1 simulation is very close to the value of the numerical $\Reyk$ at grid resolution $\Nres = 288$, \citep[see Appendix C. in ][which shows that the numerical $\Reyk \approx 2 \Nres^{4/3}\approx 3800$ for \mbox{$\Nres = 288$}]{mckee2020magnetic}. In \autoref{tab:converge_params} we also find that the characteristic resolution ($\Nc$) of $\knu$, $\keta$ and $\kp$ exceeds the resolution of this simulation. Therefore, in \autoref{fig:kurtosis} we grey the measurements from this simulation to highlight that the effects of numerical viscosity will influence the velocity gradients we measure.
} (see \autoref{tab:simulations_setup}) at $\Nres = 288$ and time-averaged over $2 \teddy$ within the kinematic dynamo regime. Note, in \autoref{eqn:kurtosis}, $\langle\dotsb\rangle_{\mathcal{V}}$ indicates the ensemble average of some quantity within the simulation volume $\mathcal{V} = L^3$. We plot the excess kurtosis, $\mathcal{K} - 3$ (offset by the kurtosis for a Gaussian distribution, which is $3$), in \autoref{fig:kurtosis}, where $\mathcal{K} = 3$ corresponds to Gaussian velocity gradients (horizontal, black-dashed line). 

We find the same (possibly universal) phenomena that \citet{schumacher2014small} reports, namely that the velocity gradient statistics transition from $\mathcal{K} - 3 \approx 0$ (Gaussian) to $\mathcal{K} - 3 > 0$ (super-Gaussian) at $\Reyk \approx 100$ (vertical, black-dashed line), in all three Cartesian directions, as is expected. Furthermore, we are able to probe simulations with lower $\Reyk$ than \citet{schumacher2014small} reported, and find sub-Gaussian ($\mathcal{K} - 3 < 0$) statistics in our lowest $\Reyk$ simulations ($\Reyk \approx 12$). We interpret this to mean that $\Reyk \approx 100$ is a transition from velocity fields having less and then more extreme dissipative events, compared with Gaussian velocity field statistics.

We note, however, that even though the transition from sub-Gaussian to super-Gaussian statistics in the velocity gradients coincides with what we call ``bonafide turbulent dynamo action'' (i.e., turbulent dynamo that conforms to the scale relations we explore in this study), a further, more detailed study would be required to explore this as a causal relation, which is beyond the scope of the present study. We also note that even though we report non-Gaussian velocity gradient statistics, our simulations have Gaussian velocity statistics for all $\Reyk$ \citep[see ][ for more details of the velocity statistics from our simulations]{federrath2013universality, seta2021saturation}.

%% DON'T CHANGE THESE LINES
\bsp % typesetting comment
\label{lastpage}
    
\end{document}

%% file: header.tex
\usepackage[T1]{fontenc}
\usepackage{ae,aecompl}
\usepackage{natbib}

\usepackage{graphicx} % Including figure files
\graphicspath{{Figures/}} % Setting the graphics path

\usepackage{amsmath}  % Advanced maths commands
\usepackage{amssymb}  % Extra maths symbols
\usepackage{multicol} % Multi-column entries in tables
\usepackage{bm}		  % Bold maths symbols, including upright Greek
\usepackage{threeparttable} % Table comments
\usepackage{float}    % for: Figure placement [H]
\usepackage{soul}     % for: Highlighting \hl{}
\usepackage{multirow} % for: Drawing horizontal lines between columns \cline
\usepackage{newtxtext, newtxmath}
\usepackage{cleveref} % Referencing a range of equations
\hypersetup{breaklinks=true}

\usepackage{hyperref}
\newcommand\aref[1]{\hyperref[#1]{Appendix~\ref*{#1}}}

\newcommand\G{\rm{G}}
\newcommand\n{\rm{n}}

\newcommand{\dd}[1]{\rm{d} #1} % derivative d
\newcommand{\pp}[1]{\partial #1} % partial derivative p
\newcommand{\driv}[2]{ \frac{\dd{#1}}{\dd{#2}} } % derivative: d{} / d{}
\newcommand{\priv}[2]{ \frac{\pp{#1}}{\pp{#2}} } % partial derivative: p{} / p{}
\newcommand{\ave}[1]{\left\langle #1 \right\rangle}
\newcommand{\rbrac}[1]{\left( #1 \right)}
\newcommand{\cbrac}[1]{\left\{ #1 \right\}}
\newcommand{\sbrac}[1]{\left[ #1 \right]}

\newcommand{\nquad}[1][1]{\hspace*{#1em}\ignorespaces}
\newcommand\Reyk{\mbox{Re}}  % kinetic Reynolds number
\newcommand\Reym{\mbox{Rm}}  % magnetic Reynolds number
\newcommand\Pranm{\mbox{Pm}} % magnetic Prandtl number, cf TeX's \Pr product
\newcommand\crit{\text{crit}}
\newcommand\turb{\text{turb}}
\newcommand\rms{\text{rms}}
\newcommand\sat{\text{sat}}
\newcommand\tkin{\text{kin}}
\newcommand\tmag{\text{mag}}
\newcommand\theory{\text{theory}}
\newcommand\teddy{t_\text{turb}}
\newcommand\Nres{N_\text{res}}
\newcommand\Nc{N_\text{c}}
\newcommand\knutheory{k_{\nu, \theory}}
\newcommand\ketatheory{k_{\eta, \theory}}
\newcommand\knu{k_\nu}
\newcommand\keta{k_\eta}
\newcommand\kp{k_p}
\newcommand\Pvel{\mathcal{P}_\tkin}
\newcommand\Pmag{\mathcal{P}_\tmag}
\newcommand\Mach{\mathcal{M}}
\newcommand\grate{\Gamma}
\newcommand\satlevel{(E_\tmag / E_\tkin)_\sat}

\defcitealias{Kolmogorov1941}{K1941}
\defcitealias{Schekochihin2002}{SBK2002}
\defcitealias{kazantsev1968enhancement}{K1968}
\defcitealias{Xu2016TurbulentGas}{XL2016}

\usepackage{scalerel,tikz}
\usetikzlibrary{svg.path}
\definecolor{orcidlogocol}{HTML}{A6CE39}
\tikzset{orcidlogo/.pic={
    \fill[orcidlogocol]
    svg{M256,128c0,70.7-57.3,128-128,128C57.3,256,0,198.7,0,128C0,57.3,57.3,0,128,0C198.7,0,256,57.3,256,128z};
    \fill[white]
    svg{M86.3,186.2H70.9V79.1h15.4v48.4V186.2z} svg{M108.9,79.1h41.6c39.6,0,57,28.3,57,53.6c0,27.5-21.5,53.6-56.8,53.6h-41.8V79.1zM124.3,172.4h24.5c34.9,0,42.9-26.5,42.9-39.7c0-21.5-13.7-39.7-43.7-39.7h-23.7V172.4z}
    svg{M88.7,56.8c0,5.5-4.5,10.1-10.1,10.1c-5.6,0-10.1-4.6-10.1-10.1c0-5.6,4.5-10.1,10.1-10.1C84.2,46.7,88.7,51.3,88.7,56.8z};
}}
\newcommand\orcidicon[1]{
    \href{https://orcid.org/#1}{\mbox{\scalerel*{
        \begin{tikzpicture}[yscale=-1,transform shape]
            \pic{orcidlogo};
        \end{tikzpicture}
    }{|}}}
}